\documentclass{aa}
\usepackage{graphicx}
\usepackage{hyperref}
\usepackage{txfonts}
\usepackage{physics}
\usepackage{siunitx}
\usepackage{float}
\usepackage[table]{xcolor}
\usepackage{amsmath}
\usepackage{orcidlink}
\usepackage{amsmath}
\usepackage[normalem]{ulem}

\newcommand{\lya}{$\mathrm{Ly}\alpha$}

\newcommand{\meanlco}{$\langle L'_\mathrm{CO}\rangle$}

\newcommand{\br}{$\mathbf{r}$}
\newcommand{\bt}{$\mathbf{t}$}
\newcommand{\s}{$\mathbf{s}$}

\begin{document}

\title{COMAP Pathfinder -- Season 2 results IV. A stack on eBOSS/DESI quasars}
   \authorrunning{D.~A.~Dunne et al.}
   \titlerunning{COMAP Pathfinder -- Season 2 results IV. A stack on eBOSS/DESI quasars}
   \author{
        D.~A.~Dunne\inst{1}\fnmsep\thanks{\email{\href{mailto:ddunne@astro.caltech.edu}{ddunne@astro.caltech.edu}}}\orcidlink{0000-0002-5223-8315}
        \and 
        K.~A.~Cleary\inst{1}\orcidlink{0000-0002-8214-8265}
        \and  J.~G.~S.~Lunde\inst{2}\orcidlink{0000-0002-7091-8779}
        \and 
        D.~T.~Chung \inst{3}\orcidlink{0000-0003-2618-6504}
        \and P.~C.~Breysse\inst{4}\orcidlink{0000-0001-8382-5275}
        \and 
        N.-O.~Stutzer\inst{2}\orcidlink{0000-0001-5301-1377}
        \and   
        J.~R.~Bond\inst{5}\fnmsep\inst{6}\fnmsep\inst{7}\orcidlink{0000-0003-2358-9949}
        \and
        H.~K.~Eriksen\inst{2}\orcidlink{0000-0003-2332-5281}
        \and
        J.~O.~Gundersen\inst{8}\orcidlink{0000-0002-7524-4355}
        \and 
        G.~A.~Hoerning\inst{9}\orcidlink{0000-0002-8677-6656}
        \and
        J.~Kim\inst{10}\orcidlink{0000-0002-4274-9373}
        \and
        E.~M.~Mansfield\inst{11}
        \and
        S.~R.~Mason\inst{8}\orcidlink{0009-0001-1267-952X}
        \and
        N.~Murray\inst{5}\fnmsep\inst{6}\fnmsep\inst{7}\orcidlink{0000-0002-8659-3729}
        \and        T.~J~Rennie\inst{12}\fnmsep\inst{9}\orcidlink{0000-0002-1667-3897}
        \and
        D.~Tolgay\inst{6}\fnmsep\inst{7}\orcidlink{0000-0002-3155-946X}
        \and
        S.~Valentine\inst{4}\orcidlink{0009-0008-0658-0321}
        \and
        I.~K.~Wehus\inst{2}\orcidlink{0000-0003-3821-7275}
        \and
        COMAP Collaboration
    }

   \institute{
       California Institute of Technology, 1200 E. California Blvd., Pasadena, CA 91125, USA % 1
       \and
       Institute of Theoretical Astrophysics, University of Oslo, P.O. Box 1029 Blindern, N-0315 Oslo, Norway % 2
       \and 
       Department of Astronomy, Cornell University, Ithaca, NY 14853, USA % 3
       \and 
       Department of Physics, Southern Methodist University, Dallas, TX 75275, USA % 4
       \and 
       Canadian Institute for Theoretical Astrophysics, University of Toronto, 60 St. George Street, Toronto, ON M5S 3H8, Canada 
       % 5
       \and
        Department of Physics, University of Toronto, 60 St.~George Street, Toronto, ON, M5S 1A7, Canada  % 6
       \and 
       David A.~Dunlap Department of Astronomy, University of Toronto, 50 St.~George Street, Toronto, ON, M5S 3H4, Canada % 7
       \and
       Department of Physics, University of Miami, 1320 Campo Sano Avenue, Coral Gables, FL 33146, USA  % 8
       \and
       Jodrell Bank Centre for Astrophysics, Department of Physics \& Astronomy, The University of Manchester, Oxford Road, Manchester, M13 9PL, U.K. % 9
       \and 
       Department of Physics, Korea Advanced Institute of Science and Technology (KAIST), 291 Daehak-ro, Yuseong-gu, Daejeon 34141, Republic of Korea %10
       \and
       Department of Physics, Cornell University, Ithaca, NY 14853, USA % 11
       \and 
       Department of Physics and Astronomy, University of British Columbia, Vancouver BC V6T 1Z1, Canada % 12
   }

   \date{Received *** / Accepted ***}

\abstract{

We present a stack of data from the second season of the CO Mapping Array Project (COMAP) Pathfinder on the positions of quasars from eBOSS and DESI. COMAP is a Line Intensity Mapping (LIM) experiment targeting dense molecular gas via CO(1--0) emission at $z\sim3$. COMAP's Season 2 represents a $3\times$ increase in map-level sensitivity over the previous Early Science data release. We do not detect any CO emission in the stack, instead finding an upper limit of $10.0\times 10^{10}\ \mathrm{K\ km\ s^{-1}\ pc^2}$ at 95\% confidence within an $\sim 18\ \mathrm{cMpc}$ box. We compare this upper limit to models of the CO emission stacked on quasars and find a tentative ($\sim 3 \sigma$) tension between the limit and the brightest stack models after accounting for a suite of additional sources of experimental attenuation and uncertainty, including quasar velocity uncertainty, pipeline signal loss, cosmic variance, and interloper emission in the LIM data. The COMAP-eBOSS/DESI stack is primarily a measurement of the CO luminosity in the quasars' wider environment and is therefore potentially subject to environmental effects such as feedback. With our current simple models of the galaxy-halo connection, we are thus unable to confidently rule out any models of cosmic CO with the stack alone. Conversely, the stack's sensitivity to these large-scale environmental effects has the potential to make it a powerful tool for galaxy formation science, once we are able to constrain the average CO luminosity via the auto power spectrum (a key goal of COMAP).

}


   \maketitle

\section{Introduction}

Despite major advances in instrumentation over the past decades, we do not yet have the observational capacity to resolve certain mysteries about the Universe's history of galaxy formation. For instance, the star formation density of the Universe is known to peak at $z\sim3$ \citep[the Epoch of Galaxy Assembly, or EoGA;][]{madaudickinson2014_sfhistory}, but the mechanism driving that peak is still uncertain. The fuel for star formation, molecular hydrogen gas (traced by molecular lines such as carbon monoxide, CO), is expensive to observe at these redshifts and galaxy surveys constraining its evolution are limited to small cosmic volumes (1-10 square arcmin) and bright galaxies \citep[e.g.][]{riechers2019_coldz, aravena2019_aspecs, lenkic2020_phibss2}. The faint end of the CO luminosity function is nearly entirely unconstrained during the EoGA, and thus empirically-rooted models for the overall CO luminosity density span several orders of magnitude \citep[][]{li2016_comodelling, padmanabhan2018_comodel, keating2020_mmime, yang2022_SAMs}. Similarly, current models of star formation require feedback from active supermassive black holes (SMBHs) to disrupt star formation in high-mass galaxies in order to match observations, but the actual mechanism and extent of this feedback is poorly understood \citep[e.g.][]{sanders1988_quasarorigins, hopkins2006_agnmodel}. Again, understanding this feedback requires tracing its effect on the fuel for future star formation in galaxies, and understanding of this feedback is limited by selection effects and small sample sizes in CO observations of active galaxies \citep[][]{hill2019, munozelgueta2022_apexqsos}.

Line Intensity Mapping (LIM) is a rapidly-maturing observational technique that measures galaxy populations as fluctuations in the line intensity field over a large 3D cosmic volume, rather than as individual catalogued objects. LIM fluctuations contain the integrated emission from all galaxies in the measured volume, even the very faintest, and thus capture the faint halos not observable by traditional means. LIM, while rapidly maturing, is a field still in its early stages, with pathfinder experiments concluded or ongoing \citep[e.g.][]{keating2015_copss1, keating2020_mmime, concerto2020_intro, cleary2021_comapoverview} and others beginning to come online \citep[e.g.][]{dore2014_spherexintro, crites2014_timeSPIE, viera2020_tim, ccatprime2023_overview}. The biggest challenge facing LIM experiments (outside of those targeting the 21\,cm line) is raw sensitivity, but they are also susceptible to many forms of systematic error (caused by, for example, instrumental effects or interloping spectral line emission).

An emerging body of work is combining LIM observations with spectroscopic galaxy catalogues to both improve sensitivity and mitigate systematic error (any major systematic error should be independent to one survey or the other). This can be done as a real-space 3D co-addition of the LIM data on the position of the objects in the galaxy survey (a `stack'), or as a real or Fourier-space cross-correlation. In their early stages, many experiments have been turning to the stack as a conceptually simpler analysis, with the potential to provide a detection in advance of the full auto power spectrum analysis \citep{keenan2021_copssstack, dunne2024_ebossstacking, chen2025_MeerKLASSstack}. A stack on LIM data probes the larger-scale distribution of emission around the galaxy survey objects in addition to emission from the surveyed objects themselves, at scales of tens of Mpc \citep{chen2025_MeerKLASSstack, dunne2025_stacktheory}. Spectroscopic galaxy catalogues which trace more massive dark matter (DM) halos (or more biased galaxies) thus yield a brighter stack, as their surroundings contain more, and brighter, galaxies. Stacks on highly biased objects such as quasars should provide the brightest possible stack emission \citep{breysse2019_agnfeedback, dunne2024_ebossstacking}.

In this work, we present a stack of LIM data from Season 2 of the CO Mapping Array Project (COMAP) Pathfinder telescope \citep{lunde2024_COMAPS2_PaperI, stutzer2024_COMAPS2_PaperII, chung2024_COMAPS2_PaperIII}, on the positions of quasars from the extended Baryon Oscillation Spectroscopic Survey \citep[eBOSS,][]{eboss_dr16} and the Dark Energy Spectroscopic Instrument (DESI) QSO survey \citep{desidr1_2025}. COMAP targets molecular gas via CO(1-0) emission around the peak of cosmic star formation ($2.4 < z < 3.4$) and eBOSS and DESI together encompass the largest survey of quasars available in this redshift range. This work follows a similar analysis done using data from COMAP Season 1 and eBOSS alone \citep{dunne2024_ebossstacking}, but this updated stack is more than $3\times$ more sensitive thanks to the improved COMAP Season 2 sensitivity. We have also considerably refined our methodology. 

We introduce the three datasets in \S\ref{sec:data}, and describe the simulated LIM and galaxy catalogue data to which we compare the stack in \S\ref{sec:simulations}. The stack methodology is described in \S\ref{sec:methods}. In \S\ref{sec:results} we present the stack results, to which we compare models in \S\ref{sec:model_comparison}. Throughout this work, we take the nine-year WMAP cosmology \citep{Hinshaw_2013} for consistency with the peak-patch simulations used in \S\ref{sec:simulations} and other COMAP works. This is a $\Lambda$CDM model with $\Omega_\mathrm{m} = 0.286$, $\Omega_\Lambda = 0.714$, $\Omega_\mathrm{b} = 0.047$, and $H_0=100h\mathrm{km\ s^{-1}\ Mpc^{-1}}$ with $h=0.7$. We assume base-10 logarithms unless otherwise stated.

\section{Data}\label{sec:data}

\subsection{COMAP}\label{sec:data:comap}
The CO Mapping Array Project (COMAP) Pathfinder instrument is a 19-feed (single-polarization) focal plane array fielded on a 10.4-m antenna \citep{lamb2021_instrument} that has been performing LIM observations targeting CO(1--0) at $2.4 < z< 3.4$ since 2019, and is still continuing to observe. Thus far, two COMAP observing seasons have taken place. Season 1 encompassed 5,200 on-sky observation hours taken over 13 months between May 2019 and August 2020 over three fields.  In this work, we used the COMAP Season 2 dataset \citep{lunde2024_COMAPS2_PaperI, stutzer2024_COMAPS2_PaperII, chung2024_COMAPS2_PaperIII}, which  added 12,300 hours of observations collected between November 2020 and November 2023, for a total of 17,500 hours. This is an increase of $340\%$ in raw data volume compared to the previous `Early Science' (ES) dataset. Combined with improvements to the data reduction pipeline that yield a much greater data efficiency \citep[described in][]{lunde2024_COMAPS2_PaperI}, COMAP Season 2 has $7\times$ more effective integration time than the ES dataset, giving a per-voxel RMS in the final maps of $25$--$50\ \mu\mathrm{K}$ \citep{lunde2024_COMAPS2_PaperI, stutzer2024_COMAPS2_PaperII}.

The COMAP time-ordered data (TOD) were processed from raw detector readout into calibrated time-domain data in brightness temperature units feed-by-feed, using a series of filters designed to reduce correlated noise and systematic errors. The COMAP calibration uncertainty is $< 2\%$, and we treated it as negligible here. The raw TOD were then binned into maps with $2'\times 2'$ spatial pixels (spaxels) and spectral channels with width $31.25\ \mathrm{MHz}$ using a simple nearest-neighbour mapmaker, as the filtered data were found to have negligible correlated noise. A map-domain PCA filter was used to remove lingering systematic effects. These steps are explained in full in \cite{lunde2024_COMAPS2_PaperI}.

Each of these steps has the potential to remove signal as well as noise -- we account for signal removed in the pipeline steps in \S\ref{sec:methodology:extraction}. The end results of the COMAP pipeline are maps that are consistent with white (Gaussian) noise.

\subsubsection{Map-level RMS Cuts}\label{sec:data:comap:map_cuts}
Because COMAP is an experiment that scans across the sky to cover its fields, the RMS noise varies considerably across each COMAP field. The outer, low-significance regions of the map contribute almost no sensitivity to the stack. We thus removed them for efficiency, and to reduce the chance of incorporating untreated sysetmatic errors from these low-significance regions into the stack. In addition to the map-level cuts used by our autocorrelation pipeline \citep{lunde2024_COMAPS2_PaperI, stutzer2024_COMAPS2_PaperII}, we mask all voxels that exceed $8\times$ the mean RMS of the 100 highest-significance voxels. We also clean out irregular regions of higher-significance voxels around the outskirts of the map (including some voxels that are totally surrounded by masked voxels). We masked all voxels which had fewer than 4 of their 8-connected neighbours unmasked. These added cuts masked $12.5\%$ more voxels than the autocorrelation masking strategy across the three COMAP fields, but increased the stack uncertainty by only $2\%$.

\begin{figure*}[ht!]
    \centering
    \includegraphics[width=0.95\linewidth]{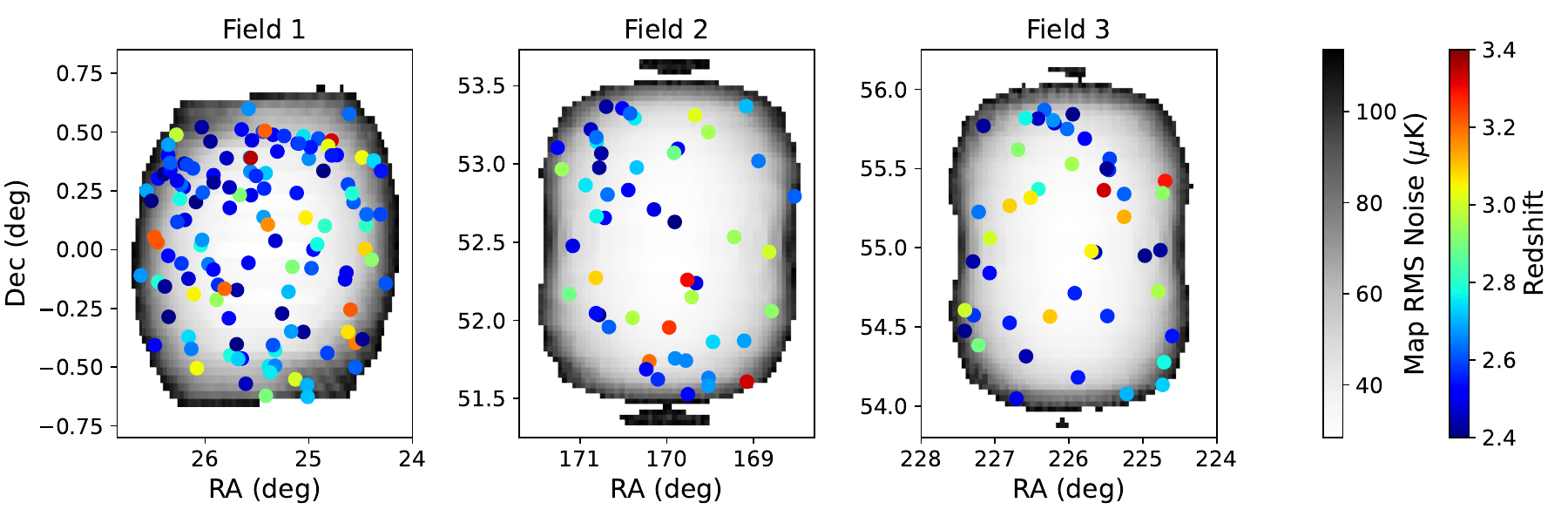}
    \caption{Spatial distribution of the eBOSS/DESI quasars compared to the footprint of the COMAP data. The COMAP footprint is coloured by the RMS in each voxel, averaged across all spectral channels, and the quasars are coloured by their redshift.}
    \label{fig:spatial_distribution}
\end{figure*}

\begin{figure}[ht!]
    \centering
    \includegraphics[width=0.95\linewidth]{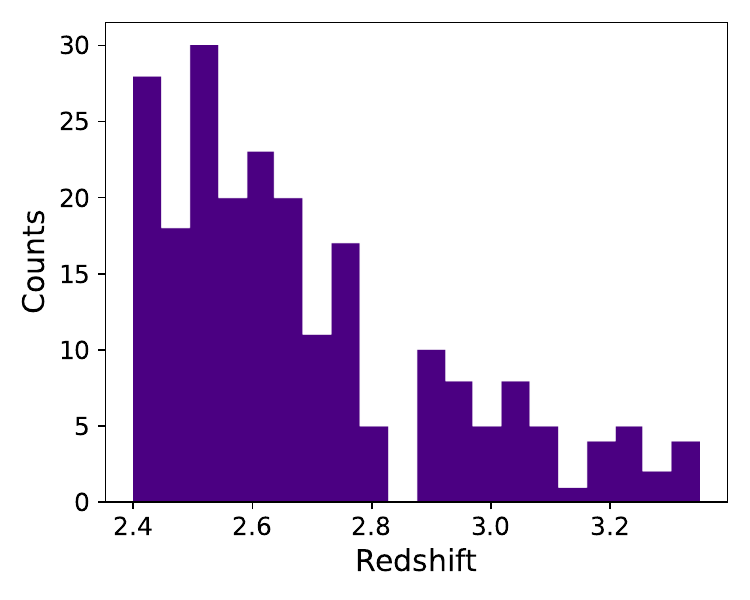}
    \caption{Redshift distribution of the eBOSS and DESI quasars included in the stack catalogue.}
    \label{fig:redshift_distribution}
\end{figure}

\subsection{eBOSS and DESI}\label{sec:data:eboss_desi}
We performed the stack on a combined catalogue of objects from the two largest spectroscopic quasar surveys available in the COMAP fields, the extended Baryonic Oscillation Spectroscopic Survey \citep[eBOSS,][]{dawson2013_sdssbossreduction, eboss_dr16, sdssdr16} , and the Dark Energy Spectroscopic Instrument \citep[DESI,][]{desidr1_2025}. Quasars are ideal targets for a stacking analysis, as they trace high-mass dark matter halos which are very highly clustered. This, more than any other factor, leads to a bright stack signal \citep{dunne2025_stacktheory}.

As we have previously performed a COMAP stacking analysis using eBOSS QSOs \citep{dunne2024_ebossstacking}, and the first season of COMAP data, we refer the reader to that work for the details of the eBOSS catalogue reduction. We took DESI data from their first Data Release \citep[DR1,][]{desidr1_2025}. DESI is still ongoing -- there will be more objects in the COMAP fields in the future as the survey continues. The DESI target selection pipeline does not account for whether a QSO was previously observed with eBOSS, so there is significant overlap between the two samples. We treated any DESI object within $1.3'$ (which is the average SDSS seeing FWHM, \citealt{delubac2017_ebosssystematics}) and $\Delta z = 0.012$ (which is the eBOSS+DESI uncertainty on the redshifts, and roughly corresponds to the $1185\,\mathrm{km\,s^{-1}}$ redshift offset we observe in quasars) of an eBOSS object as a duplicate.
This is a similar cut to that of a comparable analysis performed by the DESI team, who treated any objects within $1'$ as a match \citep[][]{desisampledefs_2024}. 
Where duplicates were present, we used the DESI position and redshift.

Additionally, we applied a systematic velocity offset of $1185\ \mathrm{km\ s^{-1}}$ to each QSO in the combined eBOSS-DESI catalogue. This value was determined in \cite{dunne2024_ebossstacking} from comparing optically-determined and CO-based redshift values for a selection of $z\sim 3$ QSOs. Astrophysically, it originates from the optical spectral lines used by eBOSS/DESI tracing bulk in- and outflows in QSOs, whereas the CO emission traces dense gas at the centre of the galaxy. We note that varying the magnitude of this offset does not meaningfully affect our results. This effect also results in a large ($\sim 1660\ \mathrm{km\ s^{-1}}$, or several COMAP spectral channels) scatter in redshift offsets, which we address in \S\ref{sec:signal_attenuation:vel_uncert}.

Between the two surveys, there are 231 QSOs in the COMAP footprint. Of these, 95 are in eBOSS alone, and 136 are either in DESI or in both surveys. The angular distribution of the galaxies across the COMAP fields is shown in Figure \ref{fig:spatial_distribution}, and their redshift distribution in Figure \ref{fig:redshift_distribution}. We note that there are fewer QSOs in this analysis than the 243 eBOSS QSOs used for COMAP stacking in \cite{dunne2024_ebossstacking}, even after the additional DESI objects are included. This is because the ES maps did not incorporate the same RMS cuts, and thus covered areas on the sky that in practice added no sensitivity. The median redshift of the quasars included in this sample is 2.62.

\section{Simulations}\label{sec:simulations}

We generated a suite of simulated LIM data cubes and associated galaxy catalogues for comparison with the real COMAP and quasar data. These were generated using the \verb|joint_limlam_mocker|\footnote{\label{note1}\url{https://github.com/delaneydunne/joint_limlam_mocker}} code described in \cite{dunne2025_stacktheory}, based on peak-patch DM halo simulations \citep[][]{bondmeyers1996_peakpatchsims, stein2019_peakpatchsims}. We refer the reader to \cite{dunne2025_stacktheory} for the methodology behind this simulation code. The suite of simulation parameters in \cite{dunne2025_stacktheory} was designed to emulate a stack of COMAP data on the positions of Lyman-$\alpha$ emitters from the Hobby-Eberly Telescope Dark Energy Experiment \citep[HETDEX,][]{hill2024_hetdexvirus2, hill2021_hetdexvirus}. In the modelling for this work, we primarily adopted the same parameters for the CO LIM map, but altered the parameters to match the quasar catalogue. We additionally tested a suite of models for CO luminosity as a function of halo mass, $L(M)$. For all other parameters, we defaulted to those used in \cite{dunne2025_stacktheory}.

\subsection{CO modelling}\label{sec:sims:co}
\begin{figure}
    \centering
    \includegraphics[width=0.95\linewidth]{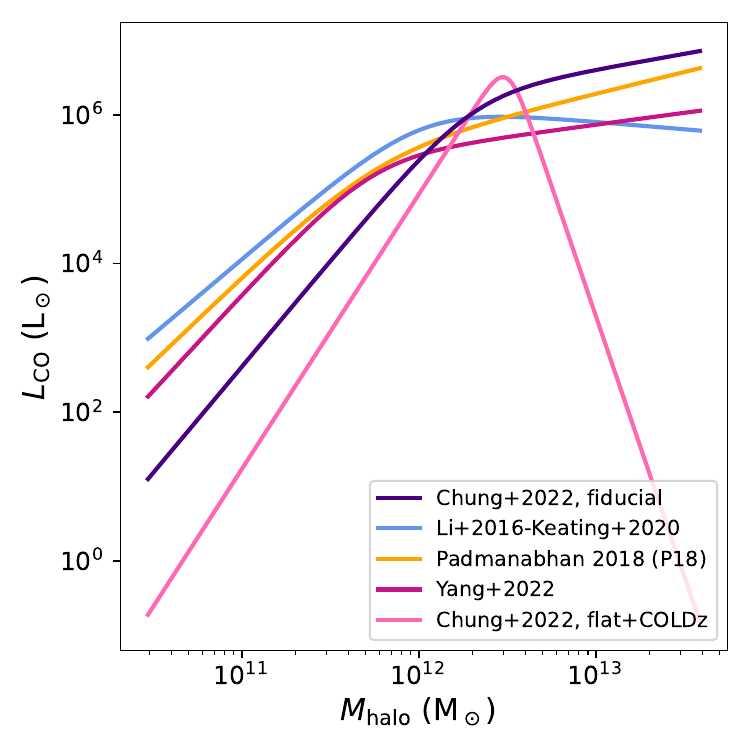}
    \caption{Models for CO luminosity as a function of halo mass used to generate simulated stacks. We defaulted to the \cite{chung2021_comapforecasts} fiducial model.}
    \label{fig:co_models}
\end{figure}

We defaulted to the `COMAP fiducial' (``UM+COLDz+COPSS") CO model from \cite{chung2021_comapforecasts}, which takes the form
\begin{equation}
    \frac{L'_\mathrm{CO}(M_\mathrm{h})}{\mathrm{K\ km\ s^{-1}\ pc^2}} = \frac{C}{(M_\mathrm{h}/M)^A + (M_\mathrm{h}/M)^B}.
\end{equation}
$A$, $B$, $C$, and $M$ are free parameters. Here, we adopted values determined using priors from \textsc{UniverseMachine} \citep{behroozi2019_universemachine}, conditioned on luminosity function constraints from the CO Luminosity Density at High Redshift survey \citep[COLDz][]{riechers2019_coldz} and the power spectrum measurement from the CO Power Spectrum Survey \citep[COPSS][]{keating2015_copss1}. The resulting parameter values were $A=-2.85$, $B=-0.42$, $\log C = 10.63$, and $\log \frac{M}{M_\odot}=12.3$. We applied a log-normal scatter with $\sigma_\mathrm{CO}=0.42\ \mathrm{dex}$. As the model space for CO at $z\sim 3$ remains extremely uncertain, we additionally tested several other models, which are shown in Figure \ref{fig:co_models}. These are 
\begin{itemize}
    \item An observationally-motivated model that determines halos' CO luminosity through a simulated star formation rate (SFR), which is in turn generated from their IR luminosities $L_\mathrm{IR}$,
    \begin{eqnarray}
        \log L'_\mathrm{CO} &=& \frac{1}{\alpha} \left[\log L_\mathrm{IR} - \beta \right], \\
        L_\mathrm{IR} &=& \frac{\mathrm{SFR}}{\delta_\mathrm{MF}}\times 10^{10}.
    \end{eqnarray}
    This is based on \cite{li2016_comodelling}. However, we took the \cite{keating2020_mmime} extension to this model, which uses the more up-to-date parameter values from \cite{kamenetzky2016_cofir} rather than \cite{carilliwalter2013_highzmoleculargas}. These values have the additional advantage of connecting the CO luminosities directly to $L_\mathrm{IR}$, without the SFR intermediary. The \cite{kamenetzky2016_cofir} parameter values are $\delta_\mathrm{MF} = 1.0$, $\alpha=1.27$, $\beta=-1.0$. We used a scatter of $0.3\ \mathrm{dex}$, and took \begin{equation}
        \frac{L_\mathrm{CO}}{L_\odot} = 4.9\times 10^{-5} \frac{L'_\mathrm{CO}}{\mathrm{K\ km\ s^{-1}\ pc^2}}.
    \end{equation}
    We will refer to this model as Li+2016-Keating+2020 going forward.

    \item A CO(1-0) model generated from fits to the Santa Cruz SAM \citep[e.g.][]{somerville2015_santacruz} from \cite{yang2022_SAMs}. This takes a redshift-dependent double-power-law form,
    \begin{equation}
        L = 2N(z)M\left[\left(\frac{M}{M_1(z)}\right)^{-\alpha(z)} + \left(\frac{M}{M_1(z)}\right)^{\beta(z)} \right]^{-1},
    \end{equation}
    with a duty cycle
    \begin{equation}
        f_\mathrm{duty} = \frac{1}{1 + \left(\frac{M}{M_2(z)} \right)^{\gamma(z)}}.
    \end{equation}
    We took values for $M_1(z)$, $N(z)$, $\alpha(z)$, and $\beta(z)$, as well as $M_2(z)$ and $\gamma(z)$ for CO(1--0) at $1.0 \leq z < 4.0$ from Table 1 of \cite{yang2022_SAMs}. We refer to this model as Yang+2022.

    \item An observationally-driven model abundance-matched onto a collection of $z=3$ CO observations by \cite{padmanabhan2018_comodel}. This is another redshift-dependent double-power-law model, 
    \begin{equation}
        L_\mathrm{CO}(M_\mathrm{h},z) = \frac{2N(z)M_\mathrm{h}}{\left(\frac{M_\mathrm{h}}{M_1(z)} \right)^{-b(z)} + \left(\frac{M_\mathrm{h}}{M_1(z)} \right)^{y(z)}},
    \end{equation}
    with $M_1(z)$, $N(z)$, $b(z)$, and $y(z)$ each given by Equation 24 of \cite{padmanabhan2018_comodel}. We also adopted the \cite{padmanabhan2018_comodel} duty cycle factor $f_\mathrm{duty}=0.1$. We refer to this model as Padmanabhan2018.

    \item The ``flat+COLDz" model from \cite{chung2021_comapforecasts}. This follows the same double power law functional form as the `fiducial' model above, but is conditioned on luminosity function constraints from the COLDz blind CO galaxy survey \citep{riechers2019_coldz} alone, with an otherwise flat prior. The coefficients that result are $A=-3.7$, $B=7.0$, $\log C = 11.1$, and $\log \frac{M}{M_\odot}=12.5$, and it has a lognormal scatter of $0.36\ \mathrm{dex}$. We refer to this model as Chung+2022, flat+COLDz. 
\end{itemize}

Overall, the COMAP fiducial model is brightest for high-mass ($M_\mathrm{h} \gtrsim 10^{12}\, \mathrm{M_\odot}$ halos, and the Li+2016-Keating+2020 model for most other halo masses. The Yang+2022 model is the faintest on average.

\subsection{Quasar modelling}\label{sec:sims:qsos}

Because of the resolution of COMAP, the stack integrates over so many DM halos (roughly 250 over $18\times18 \times 9$ cMpc) that the host galaxies of the actual quasars included in the catalogue provide only minor contributions to the total stack luminosity \citep{dunne2025_stacktheory}. Conversely, this means that the stack is extremely sensitive to the number and CO luminosity of DM halos in the vicinity of the quasars. A successful mock catalogue, then, will recover the relationship between the (observed) quasars and the underlying large-scale structure, with the CO properties of the quasars themselves only a secondary effect. 

\begin{figure}
    \centering
    \includegraphics[width=0.95\linewidth]{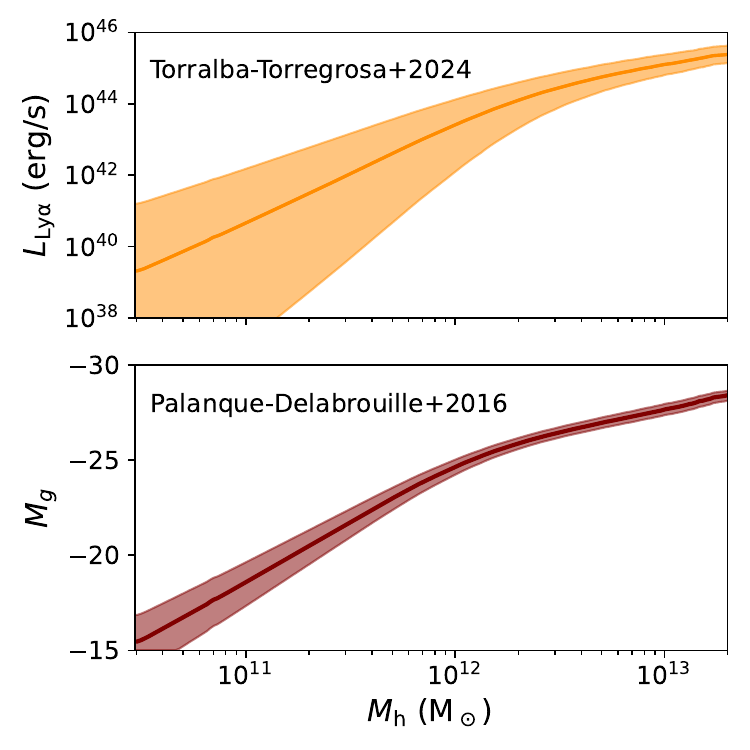}
    \caption{Models for quasar luminosity used to generate simulated catalogues, created from abundance-matching onto observed luminosity functions. The top panel shows the PAU model from \citep{torralba-torregrosa2024_quasarlyalumfunc} and the bottom panel shows the \cite{palanquedelabrouille2016_ebosslumfunc}. We defaulted to the \cite{torralba-torregrosa2024_quasarlyalumfunc} model.}
    \label{fig:quasar}
\end{figure}

As the quasar luminosity function (QLF) has considerably more observational backing from large surveys $z\sim 3$ than the CO models \citep[e.g.][]{eboss_dr16, palanquedelabrouille2016_ebosslumfunc}, we followed \cite{dunne2025_stacktheory} in generating modelled quasar luminosities by abundance-matching onto observed QLFs. We show the parametrizations of the QLF we use in this work in Figure \ref{fig:quasar}. The default model we used is based on a \lya\ luminosity function generated from quasars in the Physics of the Accelerating Universe survey \citep[PAU][]{torralba-torregrosa2024_quasarlyalumfunc}, and follows a \cite{schechter1976} form,
\begin{equation}
    \frac{dn}{dL} = \frac{\phi^*}{L^*} \left(\frac{L}{L^*}\right)^\alpha e^{-L/L^*}.
\end{equation}
We took the lowest-redshift bin, which is centred slightly above the $z=2.62$ median redshift of our combined eBOSS-DESI catalogue, yielding $\log \left( L^*/\mathrm{erg\ s^{-1}} \right)=44.89$, $\log \left(\phi^*/\mathrm{Mpc^{-3}} \right)=-6.48 $, and $\alpha = -1.44$. Through propagating forward the uncertainty on these best-fit values, we found an average scatter at the high-mass end ($M_\mathrm{halo} > 5\times10^{12} \ \mathrm{M_\odot}$) of $\sigma_\mathrm{QSO} = 0.39\ \mathrm{dex}$. We took the limiting luminosity $L_\mathrm{cut} = 10^{44}\ \mathrm{erg\ s^{-1}}$ \citep[][]{torralba-torregrosa2024_quasarlyalumfunc}. 

Additionally, we tested the QLF from \cite{palanquedelabrouille2016_ebosslumfunc}, generated from eBOSS and used for target selection for DESI. The luminosities treated in this work are absolute $g$-band magnitudes, normalized to $z=2$. The luminosity function follows a double-power law form, 
\begin{equation}
    \Phi(M_g, z) = \frac{\Phi^*}{10^{0.4(\alpha + 1)(M_g - M_g^*)} + 10^{0.4(\beta+1)(M_g - M_g^*)}}
\end{equation}
with several redshift-dependent modifications modulating $\Phi^*$, $M_g^*$, and $\alpha$. We followed these modifications, using the `luminosity and density evolution' (LEDE) functional form that the authors suggest for $z>2.2$. These modifications are
\begin{equation}
    \log\left[\Phi^* (z) \right] = \log \left[\Phi^* (z_\mathrm{p}) \right] + c_{1a} (z-z_\mathrm{p}) + c_{1b} (z - z_\mathrm{p})^2,
\end{equation}
\begin{equation}
    M^*_g (z) = M_g^* (z_\mathrm{p}) + c_2 (z-z_\mathrm{p}),
\end{equation}
and 
\begin{equation}
    \alpha(z) = \alpha(z_\mathrm{p}) + c_3 (z-z_\mathrm{p}).
\end{equation}
In this high-redshift limit, there are 8 fitted parameters ($M_g^*(z_\mathrm{p})$, $\Phi^* (z_\mathrm{p})$, $\alpha$, $\beta$, $c_{1a}$, $c_{1b}$, $c_2$, and $c_3$). We took the best-fit values for each from Table 4 of \cite{palanquedelabrouille2016_ebosslumfunc}. We calculated the luminosity function at $z=2.62$ (the median redshift of the combined eBOSS-DESI catalogue), and used this across the entire COMAP redshift range. We determined the log-normal scatter in this relation from the uncertainty on the best-fit parameters, propagating them into the $L(M)$ relation and taking the median scatter at the high-mass end ($M_\mathrm{halo} > 5\times 10^{12}\ \mathrm{M_\odot}$). This yields a scatter of 0.11 dex. 

The actual magnitude limits of the DESI quasar catalogue are defined in the $r$-band. We used the \cite{chaussidon2023_desiselection} definition of a (relative) magnitude limit of $r=23.0$, and corrected this to the $g$-band using the BOSS composite spectrum of \cite{harris2016_quasarcompositespectrum}, which provides a $g-r$ value of $0.08$ at our median redshift of $2.62$. The correction stays below $g-r = 0.2$ over our entire redshift range. We thus defined a relative magnitude limit of $g < 23.08$ for this model.
% We calculate relative magnitudes from the normalized absolute $g$-band magnitudes following \cite{palanquedelabrouille2016_ebosslumfunc}, 
% \begin{equation}
%     M_g^*(z-2) = g_\mathrm{dered} - d_M(z) - \left[K(z) - K(z=2) \right]
% \end{equation}

\section{Stacking methodology}\label{sec:methods}
The stack is a coaddition of the COMAP data at the position of each quasar. To accomplish this we primarily followed the methodology established in \cite{dunne2024_ebossstacking}, and we refer the reader to that work for the full details. Here, we briefly summarize the methodology and describe the modifications we made to account for pipeline signal loss.

\subsection{Coaddition}\label{sec:methodology:stacking}

We first converted the three-dimensional COMAP maps for each field from brightness temperature ($T_\mathrm{b}$) units into CO line luminosity values ($L'_\mathrm{CO}$) following \cite{Solomon_1997},
\begin{equation}
    L'_\mathrm{CO} = \frac{c^2}{2k_\mathrm{B}}\left(S \Delta v \right) \frac{D_\mathrm{L}^2}{\nu^2_\mathrm{obs} \left(1+z \right)^3},
\end{equation}
where
\begin{equation}
    \left(S \Delta v \right) = \frac{2\nu^2 k_\mathrm{B} T_\mathrm{b}}{c^2} \Omega_\mathrm{vox} \Delta v.
\end{equation}
Here, $\left(S \Delta v \right)$ is the integrated CO line flux in the voxel, $D_\mathrm{L}$ is the luminosity distance of the catalogue object, $\nu_\mathrm{obs}$ is the observed frequency and $z$ is the redshift. $\Omega_\mathrm{vox}$ refers to the solid angle of the sky subtended by the voxel. We then extracted from the COMAP data cubes a 3D `cubelet' centred around the spatial position and observed CO(1-0) frequency of each catalogue object. These cubelets were $31\times31\times81$ voxels ($62'\times 62' \times 2.53\ \mathrm{GHz}$, or $113\ \mathrm{cMpc} \times 113\ \mathrm{cMpc} \times 337\ \mathrm{cMpc}$), with the catalogue object itself located within the central voxel of the cubelet. For context, the full COMAP cubes are $120\times120\times256$ voxels. The cubelets were purposefully much larger than the expected extent of the stack signal in order to account for larger-scale structure.

Finally, we combined the COMAP data cubelets associated with each catalogue object voxel-by-voxel, weighting by the voxel RMS using an inverse-variance scheme to account for the inhomogeneous noise response across the COMAP data cubes. We did this separately for each field. We thus obtained a single stacked cubelet, containing the average three-dimensional profile of the CO emission surrounding the catalogued objects, for each field. We will refer to this average cubelet as $\mathbf{s}$. For presentation purposes, we summed the averaged cubelet across its central $5\times 5$ spatial voxels or $3$ spectral voxels to generate 1D spectra or 2D images, respectively. These are shown in Figure \ref{fig:stack_combim}.

\subsection{Extraction of line luminosity}\label{sec:methodology:extraction}

The goal output value of the stack is the average CO luminosity of the quasars in the eBOSS/DESI catalogue and their surroundings over some given cosmic volume, \meanlco. In \cite{dunne2024_ebossstacking}, we obtained this value by integrating the stacked cubelet data $\mathbf{s}$ over a central aperture $3\times3$ spatial pixels and $7$ spectral channels wide. However, the COMAP data reduction pipeline, intended to take the data from raw time-ordered data (TOD) to a map free of systematic errors, will act on the signal as well as the noise, and modify both the absolute magnitude and 3D distribution of the input ground-truth signal. In order to account for this, we changed the methodology by which we extract the \meanlco\ from the stacked datacube. We determined \meanlco\ by scaling a 3D model $\mathbf{r}$, containing our best approximation of the predicted stack signal, by an amplitude $a$ fit to the real stacked data $\mathbf{s}$. We then integrated over the scaled model to yield \meanlco. We did this separately for each field, as the different on-sky locations of the fields meant that the pipeline normalization filter, which applies to the TOD, has a different behaviour in RA and Dec for each field.

\subsubsection{Extraction methodology}\label{sec:methodology:extraction:method}

We extracted the signal in four steps, performing each step individually for each COMAP field. Firstly, we generated a 3D model of the signal to be stacked using simulations generated with \verb|joint_limlam_mocker|. Then, we processed the raw simulated signal \br\ by the COMAP analysis pipeline, to ensure the effects of the filters on the real data are accounted for in the model, by injecting the simulated signal into subsets of the  COMAP TOD and passing these injection simulations through the pipeline. We generated stacks of the simulated signal, both before and after processing by the pipeline. We refer to the resulting unprocessed model as \br\ and the pipeline-processed model as \bt\ (these are discussed in more detail below in \S\ref{sec:methodology:injections:templates}, and shown in Figure \ref{fig:transfer_function_inputs_mask}). Thirdly, we used the processed template \bt\ as a matched filter, and determined the amplitude $a$ that would scale it to the real stacked datacube \s,
\begin{equation}
    a = (\mathbf{t}^\mathrm{T} \mathbf{t})^{-1} \mathbf{t}^\mathrm{T} \mathbf{s}.
\end{equation}
We then determined the input stacked signal best-explained by the actual data by multiplying $a$ into the original, unprocessed stack model \br. This step also corrected for signal loss incurred by the pipeline filters. In the final step, we calculated the corrected output stack luminosity \meanlco\ by integrating over the scaled template signal $a\mathbf{r}$ in three dimensions,
\begin{equation}
    \langle L'_\mathrm{CO} \rangle = a\sum^{+1}_{k=-1}\sum^{+2}_{i,j=-2} r_{i,j,k},
\end{equation}
where $r$ is the voxel of the model cube $\mathbf{r}$ indexed by $i$ and $j$ in the two spatial dimensions and $k$ in the spectral (\br\ is in units of line luminosity, so the area and spectral width of each voxel are accounted for). In this work, we integrated over the central $5\times 5$ spaxels (in contrast to the $3\times 3$ used in \citealt{dunne2024_ebossstacking}) because the actual stack signal is dominated by cosmological clustering and is thus considerably more broadened than a point source. Additionally, we summed over $3$ spectral channels, rather than $7$, and accounted for signal spread out of this central aperture due to uncertainties on the systematic redshifts of the quasars using a scaling factor $f_\mathrm{VI}$ (see \S\ref{sec:signal_attenuation:vel_uncert}). 

This is a simplified version of a full `adaptive matched filtering' analysis which we will present in \cite{mansfield2025_adaptivephot}. There are two advantages to this strategy: extracting \meanlco\ using a template weights most heavily the voxels which contain the most signal, and it enables us to efficiently account for signal loss due to the analysis pipeline. 
Finally, this methodology does introduce a model-dependence to the output \meanlco. We describe how we accounted for this in \S\ref{sec:error_budget:model_dep}.

\subsubsection{Template generation}\label{sec:methodology:injections:templates}

In order to perform the matched filtering correctly, we needed to use as accurate an estimation of the raw signal output from the COMAP stack as possible. Thus, the templates for each field needed to incorporate any modification of the signal incurred during signal processing. To generate an accurate template $\mathbf{t}$, we thus applied the pipeline filters to the simulated signal. We did this by injecting simulated signal into a subset of the raw COMAP time-ordered data (TOD), and passing it through the full pipeline and mapmaker. Because we are not detecting CO signal in either auto-correlation or this stacking analysis, and because the simulated halo distribution will not line up with the real one in any case, the raw data functioned as noise (we verify this in Appendix \ref{app:signal_injection_boost}). Using signal injected into raw data allowed us to estimate signal loss using the actual COMAP noise distribution, thus including its angular and frequency shape and any low-level systematics.

We created 100 different realizations of a simulated CO fluctuation cube and associated QSO catalogue using the \verb|joint_limlam_mocker| code and default parameters described in \S\ref{sec:simulations}. We used 50 different peak-patch realizations, generating luminosity values using two different random seeds for each (for a total of 100 realizations). This is increased from the 10 unique realizations used by \cite{lunde2024_COMAPS2_PaperI} for the auto-correlation analysis, in order to account more thoroughly for the effects of cosmic variance (see \S\ref{sec:error_budget:sample_variance}). We then injected each simulation realization into an unique subset of the full TOD (using a hundredth of the total data volume) for COMAP Fields 1 and 2, boosting the injected signal by a factor of 50. This boost had to be chosen carefully -- we required a high-significance template for effective scaling, but the signal still needed to be small enough to not cause non-linearities in any of the pipeline steps. We discuss our choice of boost in more detail in Appendix \ref{app:signal_injection_boost}. Each of the 100 TOD subsets with injected signal was passed through the full COMAP analysis pipeline and mapmaker, resulting in 100 independent mock LIM cubes for each field, each with accurate on-sky footprints and filtering. 

\begin{figure*}
    \centering
    \includegraphics[width=0.99\linewidth]{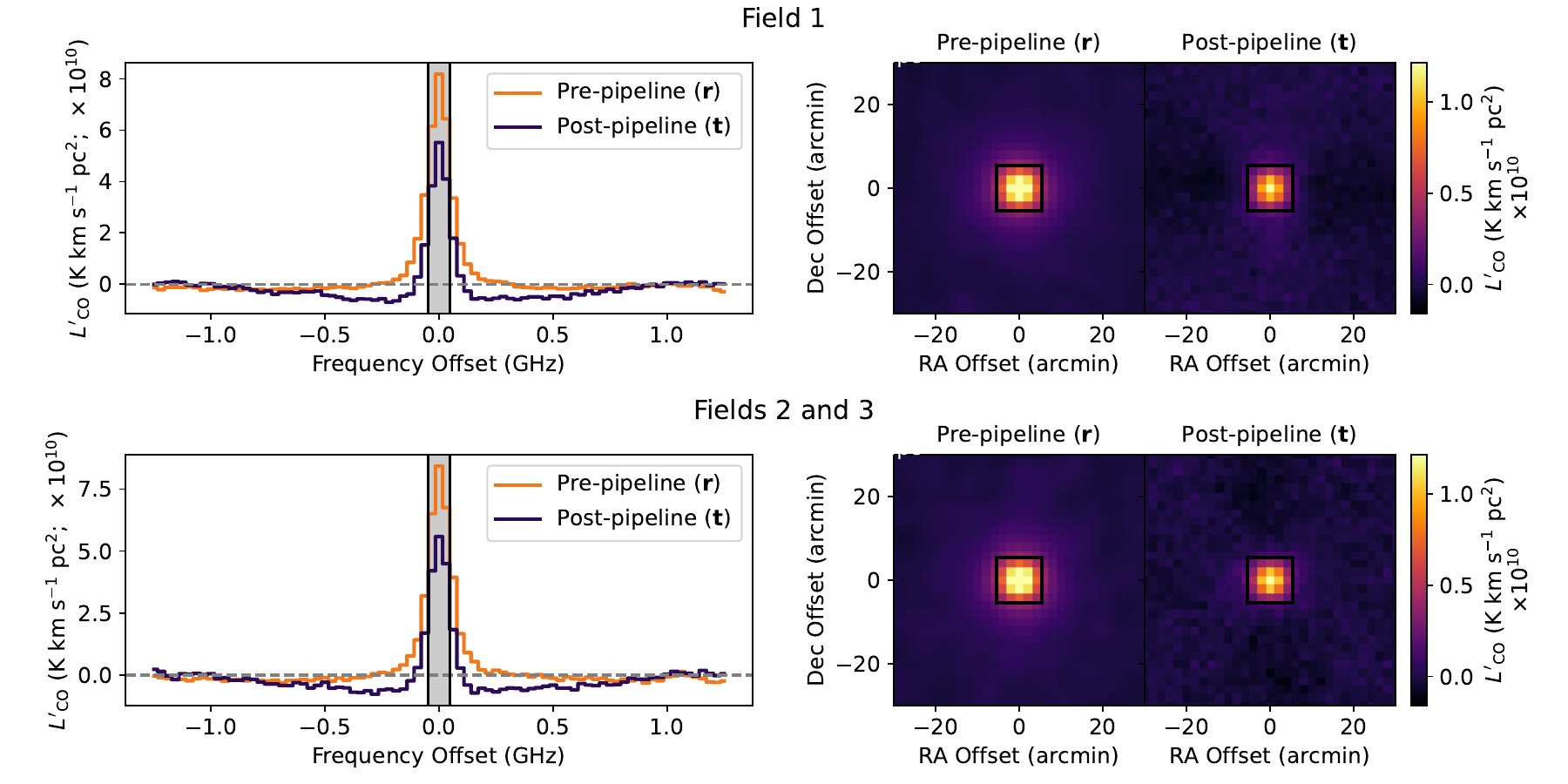}
    \caption{Averaged simulated stacks used to account for pipeline signal loss in the stack, shown here for COMAP Field 1 above and Field 2 below. We show the stacked spectra on the left (averaging across the central $5\times 5$ spaxels for presentation purposes) and the stacked images (averaging across the central three spectral channels for presentation purposes) on the left. The apertures over which we sum are shown in black. The spatial and spectral effects of the COMAP analysis pipeline, and their different effects for each field, are visible.}
    \label{fig:transfer_function_inputs_mask}
\end{figure*}

For each field of each injected realization, we then performed a mock stack. For efficiency, we used the template generated for Field 2 for both Field 2 and Field 3, as both fields traverse the sky in a similar fashion, making their scan pattern almost identical in Equatorial coordinates. This produces very similar pipeline effects, and they are thus expected to have very similar templates. We reproduced the quasar density in the real COMAP-eBOSS/DESI stack by stacking on the same number of mock quasars as are present in each real field (the injections recover the same hit patterns as the real COMAP fields, so the cosmic volumes covered by each field were also accurate). We rounded the number of quasars in Fields 2 and 3 to 50 (From 53 and 47). We simulated the uncertainty in the quasar systemic redshifts by offsetting the spectral position of each mock quasar by a random value pulled from a normal distribution with standard deviation $\sigma_v = 1660\ \mathrm{km\ s^{-1}}$ (\S\ref{sec:data:eboss_desi}). This resulted in a 3D stacked cubelet for each realization. The 131-object Field 1 stacks with a $50\times$ signal boost had S/N ratios of 6 on average in the central voxels, and the 50-object stacks for Field 2 and 3 had S/N ratios of 4, calculated using the map-level RMS noise propagated from the TOD. Finally, we averaged together the stacked cubelets for each of the 100 different realizations, voxel-by-voxel, using inverse-variance weighting. The resulting averaged cubelet is our template $\mathbf{t}$, shown in Figure \ref{fig:transfer_function_inputs_mask}. When averaged over 100 iterations, the S/N ratio of the central voxel of the template rises to $67$ for Field 1 and $49$ for Fields 2 and 3.

\subsection{Variance}\label{sec:methods:variance}

While it's possible to determine the uncertainties in each stack from propagating through the RMS noise from the TOD, this methodology doesn't capture any potential residual correlated noise in the maps, or any of the noise processing by the filters in the pipeline. Thus, in order to robustly calculate the variance in the output stacks, we bootstrapped stacks on randomized catalogues and measured the distribution of their measured line luminosity values.  We randomized the catalogues using the same method as was used in \cite{dunne2024_ebossstacking}, offsetting the coordinates of each object in the stack catalogue by some value (between 1 and 10 spaxels in the angular directions, and between 1 and 10 channels in the spectral direction. This works well for the eBOSS/DESI catalogues, as both are fairly spatially homogeneous. We ran the stack 10,000 times, each time using a different random catalogue realization generated using this method. We calculated \meanlco\ from the resulting stacked cubelets using the methodology outlined in \S\ref{sec:methodology:extraction}. We then fit a Gaussian distribution to the output \meanlco\ values and took the standard deviation of this Gaussian to be the uncertainty on the stack. This was done separately for each field, because both the average RMS noise and the number of quasars included in the stack varies between fields. 

As demonstrated in \cite{chen2025_MeerKLASSstack}, this bootstrapped strategy only accounts for variance due to instrumental noise (both correlated and uncorrelated), and does not account for sample variance, which also affects the stack considerably. However, sample variance only applies to actual signal, and in this work we achieve only an upper limit, so the sample variance does not affect the limit directly. Instead, we counted sample variance as an uncertainty on results output from our modelling; we discuss this in \S\ref{sec:error_budget:sample_variance}.

\section{Results}\label{sec:results}

\begin{figure*}
    \centering
    \includegraphics[width=0.99\linewidth]{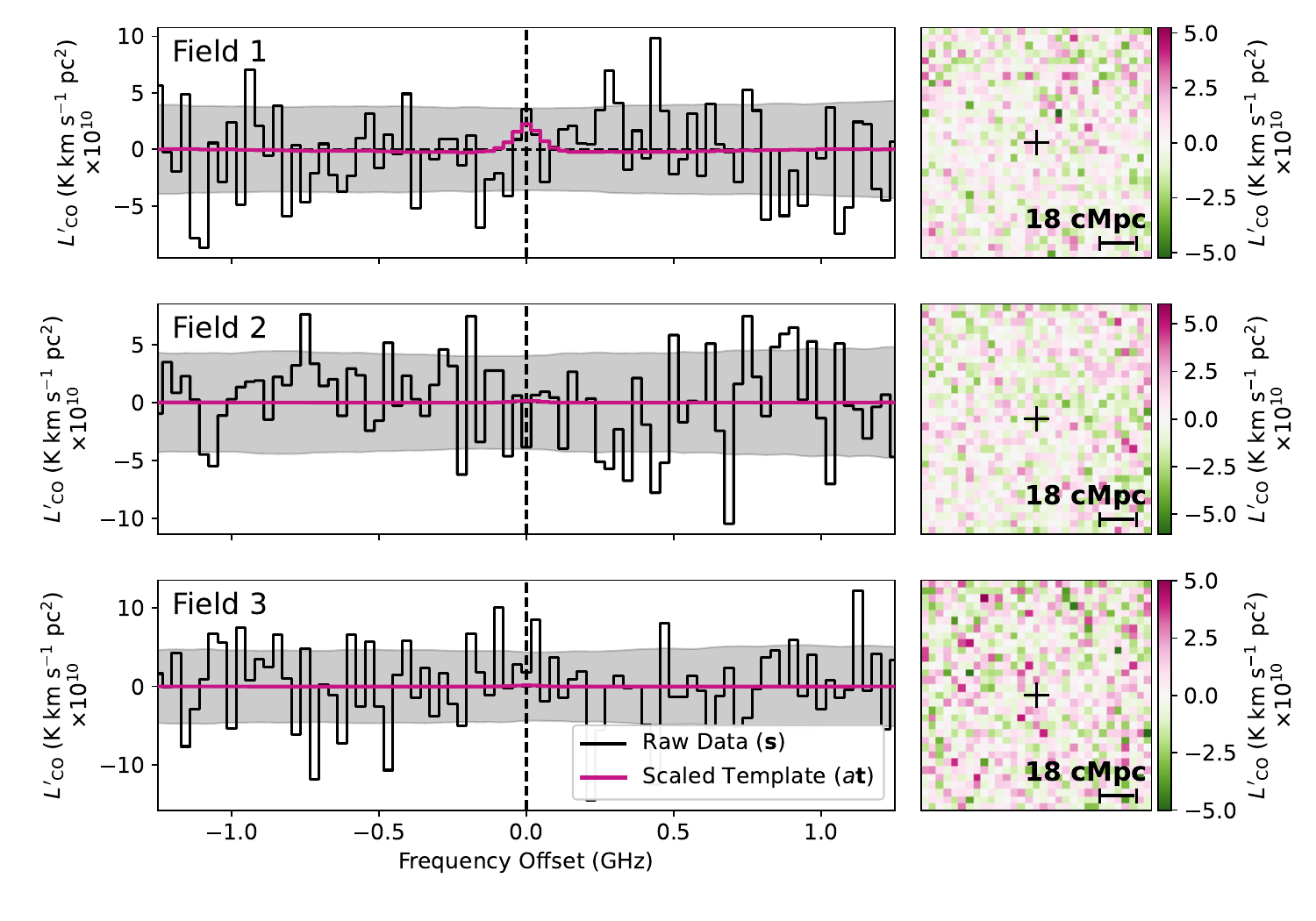}
    \caption{Stacked spectra (left) and 2D images (right) for COMAP data in each of the three COMAP fields at the positions of the eBOSS/DESI QSOs. The per-channel RMS values are shown as grey shaded regions in the stacked spectrum, and the scaled template $a\mathbf{t}$ is overlaid. The stack is not corrected for any pipeline signal-loss effects, and the templates have the same signal loss present.}
    \label{fig:stack_combim}
\end{figure*}

We show the final COMAP-eBOSS/DESI stacks for each field in Figure \ref{fig:stack_combim}, noting that the extended 2D image and 1D spectrum representations of the stack are uncorrected for any pipeline signal loss effects. We calculated the scale factor $a$ for each field, using its respective template \bt, finding $a = [0.406, 0.026, 0.027]$ for Fields [1,2,3] respectively. The scaled templates $a\mathrm{t}$ are also shown in Figure \ref{fig:stack_combim}. We then determined \meanlco\ for each field using its respective model \br, finding $\langle L'_\mathrm{CO}\rangle = [8.44, 0.56, 0.59]\times 10^{10}\ \mathrm{K\ km\ s^{-1}\ pc^2}$ for Fields [1,2,3].

\begin{figure}
    \centering
    \includegraphics[width=0.95\linewidth]{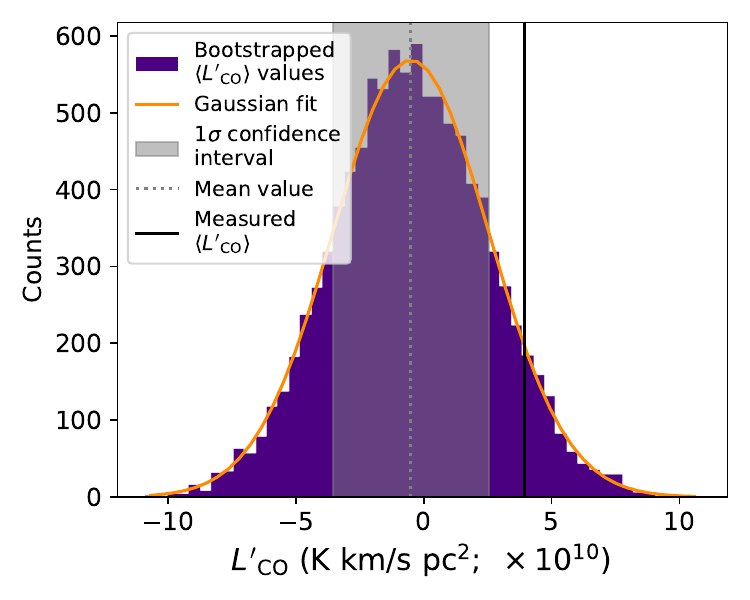}
    \caption{ Bootstrapped \meanlco\ values determined from stacks on randomized eBOSS/DESI catalogues, with a Gaussian fit. The mean and $1\sigma$ confidence interval of the Gaussian are shown in grey. }
    \label{fig:stack_bootstrap}
\end{figure}

We calculated the variance for each field using bootstrapping, as described in \S\ref{sec:methods:variance}. We combined the values extracted for each field using inverse-variance weighting, using the outputs from the bootstraps in each field to determine the variances. We found an overall output stack luminosity for the COMAP-eBOSS/DESI stack of $\langle L'_\mathrm{CO}\rangle = 4.15 \times 10^{10}\ \mathrm{K\ km\ s^{-1}\ pc^2}$, with an uncertainty $\sigma_{\langle L'_\mathrm{CO}\rangle} = 2.94 \times 10^{10}\ \mathrm{K\ km\ s^{-1}\ pc^2}$. The stack therefore has a significance of $1.41\sigma$, and is a non-detection. The $2\sigma$ (95\%) upper limit is $\langle L'_\mathrm{CO}\rangle \leq 10.0\times 10^{10}\ \mathrm{K\ km\ s^{-1}\ pc^2}$. A histogram showing the bootstrapped overall output luminosity values is shown in Figure \ref{fig:stack_bootstrap}, along with fits to a Gaussian distribution. Averaging the scale factors for each field together using the same inverse-variance weighting, we found an overall scale factor for the stack of $a=0.20$. Physically, this means that the scaling of the fiducial model that best explains the data has a fifth of the overall luminosity of our fiducial model. 

\section{Comparison to models}\label{sec:model_comparison}

To put this upper limit value into context, we compared it directly to models for CO emission stacked on QSO positions at $z\sim 3$. While these models innately include many factors which could affect the output signal (such as astrophysical line broadening of CO, instrumental beam smearing, and the relationship to the large-scale structure (LSS) of both the CO emission and the QSOs), there remain several experimental and modelling factors that could serve to either attenuate the stacked signal or add variance, and thus must be accounted for when comparing the real stack luminosity to modelled versions. These are in addition to signal loss from the COMAP analysis pipeline, which is accounted for when extracting \meanlco. We discuss these factors below. The effects of each on the simulated signal are shown in Figure \ref{fig:numbers_after_corrections}.

\subsection{Modelled luminosity values}\label{sec:model_comparison:values}
\begin{figure}
    \centering
    \includegraphics[width=0.98\linewidth]{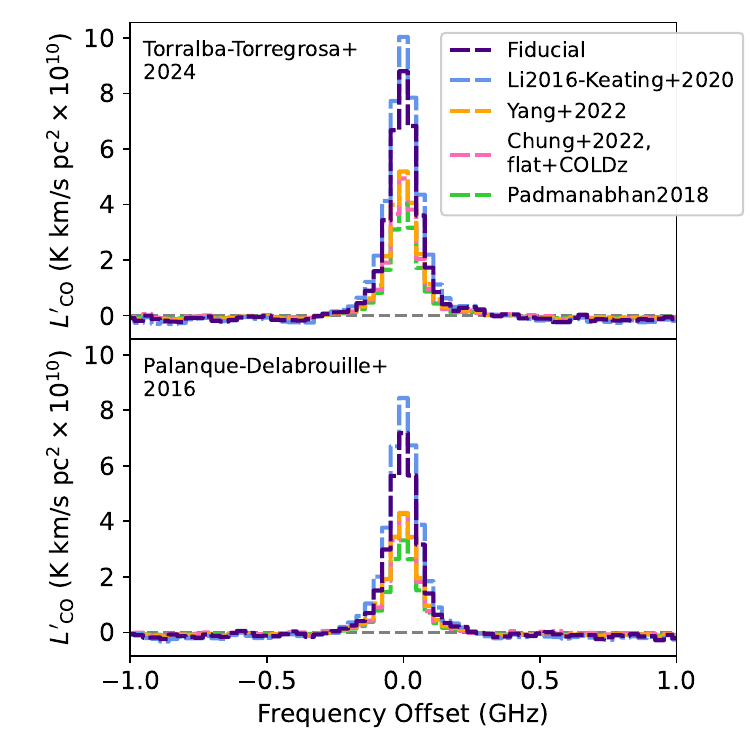}
    \caption{Spectra for each CO model choice, shown for a quasar catalogue generated using the \lya-based Torralba-Torregrosa+2024 model (top) and the eBOSS g-band luminosity function Palanque-Delabrouille+2016 (bottom). Spectra are shown before correcting for any of the attenuation factors discussed in \S\ref{sec:signal_attenuation}.}
    \label{fig:model_spectra}
\end{figure}

We tested the five CO models and two QLF models listed in \S\ref{sec:sims:co}. In each case, we determined the modelled \meanlco\ by averaging 50 simulated stack realizations. Each simulated stack consisted of three independent mock COMAP fields (simulated using one of 25 peak-patch realizations each with two random seeds for determining luminosity values), each with the real COMAP hit pattern and masking for that field. Unlike the injection simulations presented in \S\ref{sec:methodology:injections:templates}, we did not pass these mock maps through the COMAP pipeline, instead keeping them as the `true' CO distribution unadulterated by the pipeline filters (i.e.~as $\mathbf{r}_\mathrm{model}$ cubes). In each field, we generated a mock catalogue cut to the same number of objects as that field in the real stack (131, 53, and 47 for Fields 1, 2, and 3), and stacked over that catalogue. This is because quasars tend to occupy more massive DM halos in a region, which are rare \citep[e.g.][]{richardson2012_quasarHOD, hopkins2008_mergersandquasars, chowdhary2025_quasarhod}. If either the cosmic volume covered by the LIM experiment or the number of catalogued objects is much altered, the mock halo selection will go farther into the lower-mass end than is realistic, affecting the output values. The \meanlco\ values for each model are listed in Table \ref{tab:model_lums}, both before any corrections for experimental sources of signal attenuation and with uncertainties from raw sample variance (\S\ref{sec:error_budget:sample_variance}), and after the corrections are made. The uncorrected spectra are shown in Figure \ref{fig:model_spectra}.

\begin{table}[]
\caption{\label{tab:model_lums} Stack \meanlco\ values for each of the CO and galaxy catalogue models we test.}
    \centering
    \begin{tabular}{ccc}
    \hline
   Model & TT24\tablefootmark{a} \meanlco & PD16\tablefootmark{b} \meanlco \\
    & ($\mathrm{K\ km\ s^{-1}\ pc^2})$ & ($\mathrm{K\ km\ s^{-1}\ pc^2})$ \\
     & $\times 10^{10}$ & $\times 10^{10}$ \\
     \hline 
     
   \multicolumn{3}{c}{\textit{Before corrections}} \\
   \hline
       Fiducial\tablefootmark{c} & $26.87 \pm 2.43$ & $22.32 \pm 1.88$ \\
       Li+2016-Keating+2020 & $30.53 \pm 2.23$ & $26.36 \pm 1.70$ \\
       Yang+2022 & $15.83 \pm 1.17$ & $13.45 \pm 0.91$ \\
       flat+COLDz\tablefootmark{c} & $14.86 \pm 1.14$ & $13.25 \pm 1.17$ \\
       Padmanabhan2018 & $12.34 \pm 0.97$ & $10.38 \pm 0.77$ \\
    \hline
   \multicolumn{3}{c}{\textit{After corrections}} \\
   \hline
       Fiducial\tablefootmark{c} & $21.19 \pm 2.15$ & $17.55 \pm 1.71$ \\
       Li+2016-Keating+2020 & $24.36 \pm 2.25$ & $20.81 \pm 1.64$ \\
       Yang+2022 & $12.59 \pm 1.15$ & $10.61 \pm 0.86$ \\
       flat+COLDz\tablefootmark{c} & $11.79 \pm 1.06$ & $10.49 \pm 1.09$ \\
       Padmanabhan2018 & $9.79 \pm 0.91$ & $8.17 \pm 0.69$ \\
    \hline
    \end{tabular}
    \tablefoot{Top values are shown before correcting for any attenuation due to experimental parameters, with uncertainties from raw sample variances. Bottom values are shown after applying all corrections to both the signal and the variance. \tablefoottext{a}{The quasar model from \cite{torralba-torregrosa2024_quasarlyalumfunc}.} \tablefoottext{b}{The quasar model from \cite{palanquedelabrouille2016_ebosslumfunc}.} \tablefoottext{c}{From \cite{chung2021_comapforecasts}.}}
\end{table}

\begin{table}[]
\caption{\label{tab:experiment_corrections} Experimental sources of signal attenuation and variance, and their contribution to the stack.}
    \centering
    \begin{tabular}{ccc}
    \hline
   Description & Factor & Value \\
   \hline
   \multicolumn{3}{c}{\textit{Signal attenuation}} \\
   \hline
    Pointing uncertainties & $f_\mathrm{PO}$ & 1.05 \\
    Redshift uncertainties & $f_\mathrm{VI}$ & 1.21 \\
    \hline
    \multicolumn{3}{c}{\textit{Variance}} \\
    \hline
    Model-dependence & $d_\mathrm{MD}$ & 1.07 \\
    LIM interloper emission & $d_\mathrm{LI}$ & 1.04 \\
    \hline 
    \end{tabular}
\end{table}

\subsection{Sources of signal attenuation}\label{sec:signal_attenuation}

As we have no signal to compare to in the data themselves, we applied any factor that could attenuate signal directly to the models. In each case, we treated the attenuation as a simple multiplicative factor $f$. The effects are shown on our fiducial CO model in Figures \ref{fig:numbers_after_corrections}.

\subsubsection{Pointing uncertainty}\label{sec:signal_attenuation:pointing}
There is a known issue with pointing in the COMAP Pathfinder telescope leading to systematic offsets between real and assumed source locations. Using a series of bright radio calibrators (including Taurus A, Cassiopeia A, and Cygnus A), these offsets were determined to be systematically $< 1'$ in both the azimuthal (az) and elevation (el) directions, with an additional scatter of $< 1'$. The az and el offsets are dependent on the absolute az-el being observed. For Field 1 (which is at $0^\circ$ declination), this translates to slight (sub-arcmin) offset in RA and Dec, with $\sim 0.8'$ scatter. For the other two fields, which are at higher declination, offsets occur in different directions when the fields are rising and setting, spreading the RA-Dec scatter along the RA axis. In each case, any systematic offset and the majority of the scatter is contained within a single COMAP spaxel. We verified this by stacking on radio continuum sources which appear in continuum-leakage maps (Appendix \ref{app:verification}), finding no systematic offset in the continuum sources and very little additional spread in the signal.

This effect can be ignored in the COMAP autocorrelation, but the stack is predicated on the ability to centre the LIM signal on regions of sky containing catalogue objects, and thus will be attenuated by any imprecision in the instrumental pointing. We determined the magnitude of this attenuation by convolving kernels modelling the RA-Dec pointing scatter described above into the maps for each field, noting that these are upper limits on the magnitude of the pointing errors. We found that the effect is indeed negligible -- signal is attenuated by a factor of 1.04 in Field 1 and by a factor of 1.07 in Fields 2 and 3, using these upper limit values. Combining these factors weighted by the number of objects in each field yields an overall attenuation of 1.05.

\subsubsection{Catalogue redshift uncertainty}\label{sec:signal_attenuation:vel_uncert}
Redshift uncertainty in the galaxy catalogue can attenuate the stacked signal by spreading luminosity outside of the central aperture where signal is expected \citep{dunne2025_stacktheory}. We adopted the magnitude of the scatter calculated in \cite{dunne2024_ebossstacking} -- a redshift uncertainty of $1669\ \mathrm{km\ s^{-1}}$ (in addition to a systematic offset, which is accounted for above). We treated the redshift uncertainty directly for each model, by repeating the stack described above (\S\ref{sec:model_comparison:values}), offsetting the velocity of each catalogue offset by a value randomly drawn from a Gaussian distribution with zero mean and standard deviation equivalent to the redshift uncertainty. The signal is attenuated on average across models by $f_\mathrm{VI} = 1.21$. Instead of simply dividing the \meanlco\ output from each model by this value when accounting for redshift uncertainty, we used the actual \meanlco\ value from the stack with the introduced redshift scatter.

\subsection{Error budget}

Certain experimental and modelling parameters serve not to attenuate potential signal directly, but instead to increase the uncertainty on either the data (beyond the instrumental noise calculated using bootstraps in \S\ref{sec:results}) or model values. Again, we treated each of these as multiplicative factors, labelling them as $d$.

\subsubsection{Model-dependence}\label{sec:error_budget:model_dep}

Our method of signal extraction makes one key assumption -- that the spatial and spectral profile of the real stacked signal matches that of the model. The stack is not a point source, and the shape can vary significantly with the CO and quasar model assumed. In \cite{dunne2025_stacktheory} we found that the stack profile is fairly well-fit by a double-Gaussian model in the spatial axes, and that the width of each can change by $5-10\%$ with the CO model used, and much more with galaxy catalogues targeting different galaxy types (up to $30\%$ between a $z\sim3$ LAE model and a $z\sim 3$ quasar model, although there would be considerably less variation between the fairly well-characterized quasar models explored here). It is thus critical to know how the output stack luminosity changes with the adopted model. 

The injection simulations are computationally expensive, prohibiting us from determining a template $\mathbf{t}$ for each CO model and applying it directly to the real data. Instead, we tested for model-dependence by scaling the pre-pipeline stack on the fiducial model $\mathbf{r}_\mathrm{fid}$ to the pre-pipeline model stack for each other model $\mathbf{r}_\mathrm{model}$, and then extracting the \meanlco\ from the scaled fiducial model $a_\mathrm{model}\mathbf{r}_\mathrm{fid}$. The output model \meanlco\ values calculated using this method differed by at most 6.9\%, with the Chung+2022 (flat+COLDz) model the most different. We account for this effect as a multiplicative factor $d_\mathrm{MD} = 1.069$ applied directly to the uncertainty on the data value. 

\subsubsection{Sample variance}\label{sec:error_budget:sample_variance}
In addition to the variance from noise in the COMAP data, which is determinable through bootstrapping and is driving the level of the upper limit we report in \S\ref{sec:results}, there is in an inherent variance in the level of the expected signal in our models. This arises from several sources, including basic cosmic variance, scatter in the CO brightnesses of DM halos, and scatter in the number and mass of DM halos in the vicinity of objects included in galaxy catalogues. This is the difference between the `shuffled covariance' and `true mock covariance' discussed by \cite{chen2025_MeerKLASSstack}. This variance applies directly to the luminosity of the signal itself, and thus is impossible to account for in our stack value, which detects no luminosity. However, it does apply to the luminosity values returned from our modelling, and will bring down the significance of any discrepancy between the models and our upper limit. 

To characterize sample variance for each of the models we test, we calculated the mean absolute deviation (MAD) between the stack \meanlco\ value returned from each of 50 independent simulation realizations. The sample variance is sensitive to the amount of cosmic volume probed by the LIM cubes, another reason for our recreation of the exact volume probed by the stack for each simulation realization (including the masking and hit patterns of the actual COMAP data). Regions with fewer hits are downweighted and contribute less to the overall stack luminosity, effectively decreasing the cosmic volume probed by the stack. The final variances are listed as the uncertainties on each model luminosity in Table \ref{tab:model_lums}. In this particular stacking configuration, these uncertainty values are of similar magnitude to those from the random noise in the COMAP stack data.

\subsubsection{LIM interloper emission}\label{sec:error_budget:LIM_interlopers}
While $z\sim 3$ CO(1--0) emission is fairly isolated in the COMAP bandwidth, interloper spectral line emission from higher- or lower-redshift ensembles of galaxies can still disrupt the CO signal. In the case of COMAP, this interloper emission will come primarily from $z\sim8$ galaxies emitting in CO(2--1). The exact degree of contamination from this emission is unknown (because CO at high redshifts is largely unconstrained), but it is estimated to be roughly 30\% of the luminosity of the $z\sim 3$ CO(1--0) in map space \citep{chung2024_globalsignals}. The interloper contamination is heavily dependent on the number of objects being stacked -- we found a roughly 20\% scatter from a 10\% interloper (on a stack on an LAE catalogue) in a 100-object stack in \cite{dunne2025_stacktheory}. We thus repeated this analysis of \cite{dunne2025_stacktheory} using the exact quasar parameters -- we performed 99 simulation realizations with a mock galaxy catalogue generated using the Torralba-Torregrosa+2024 QLF model and 10\% LIM interloper contamination, taking 231-object stacks on each. We found that the LIM interloper contamination should add 3.8\% scatter. This number is considerably less than the scatter from comparable 100-object stacks in \cite{dunne2025_stacktheory}, due to a combination of the brighter signal from the quasar stack and the smaller aperture being summed over. We thus took $d_\mathrm{LI} = 1.038$.

\begin{figure}[t!]
    \centering
    \includegraphics[width=0.95\linewidth]{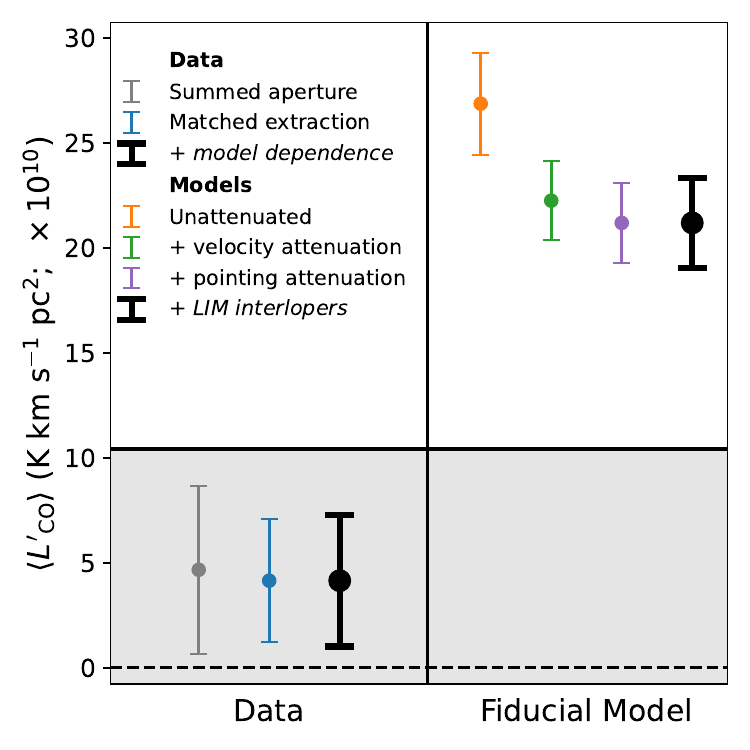}
    \caption{Factors attenuating signal or increasing variance, shown for the upper limit from the COMAP-eBOSS/DESI stack and the fiducial model from \cite{chung2021_comapforecasts}. The corrections applied to the other models are qualitatively similar. The shaded region indicates the $2\sigma$ upper limit from the data alone. The various corrections are outlined in \S\ref{sec:model_comparison}.}
    \label{fig:numbers_after_corrections}
\end{figure}

\section{Interpretations}\label{sec:discussion}
\begin{figure}
    \centering
    \includegraphics[width=0.95\linewidth]{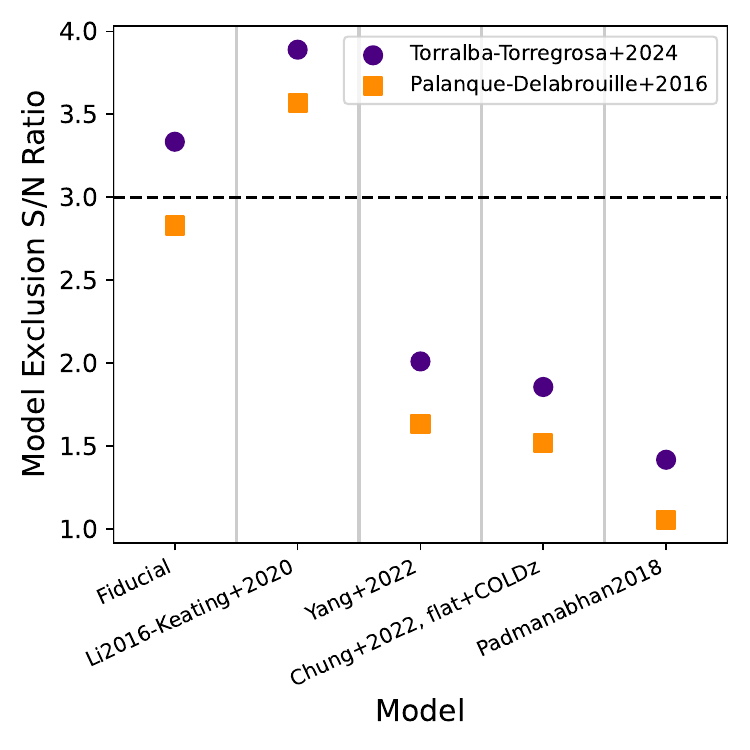}
    \caption{S/N ratio at which models are discrepant with the COMAP-eBOSS/DESI upper limit for various models of CO emission, including all potential sources of uncertainty and signal attenuation enumerated in \S\ref{sec:model_comparison}.}
    \label{fig:model_snrs}
\end{figure}

In this work, we present stacks of CO emission from COMAP on the positions of quasars from eBOSS and DESI. The stack is a biased tracer of the overall CO luminosity, probing (in this case) the average CO luminosity of an $18 \times 18\times 9\,\mathrm{cMpc}$ region around surveyed quasars. We do not detect any cosmic CO in the stack, and the upper limit we place on the stacked signal falls considerably below the modelled stack luminosity for simulations incorporating several different CO and quasar models. We have attempted a comprehensive census of possible sources of signal loss and uncertainty (Figure \ref{fig:numbers_after_corrections}), which brings the modelled luminosity values closer to the limit placed by the COMAP-eBOSS/DESI stack, but still leaves a $3.3\sigma$ disagreement between the two using our fiducial modelling parameters (the `UM+COLDz+COPSS' model from \citealt{chung2021_comapforecasts} and the Torralba-Torregrosa+2024 quasar model). We show the disagreement as a simple S/N ratio for this model and our suite of other models in Figure \ref{fig:model_snrs}. Stacks on the Li+2016-Keating+2020 CO model are discrepant with our upper limit to $> 3\sigma$ for both choices of quasar luminosity function.

Because the stack is not a straightforward tracer of cosmic CO emission, interpretation of this result is somewhat complicated. The stack does not directly probe the average CO emission in the universe as a whole. It also does not directly probe the average CO emission from the galaxies hosting the quasars --  in general the contribution of the objects surrounding the catalogued objects to the stacked luminosity is far greater than the stacked luminosity of the objects themselves \citep{dunne2025_stacktheory}. For higher-mass DM halos such as those containing quasars this difference is not as marked as it is for lower-mass halos (roughly a factor of 6 for our fiducial modelling), but the host galaxies are still a minor factor. For the purposes of discussion, we will also assume that all experimental considerations affecting \meanlco\ have been accounted for in \S\ref{sec:model_comparison}, although there is always a possibility that our result could be indicative of an unaccounted-for experimental systematic error. For discussion purposes, we break the remaining potential interpretations into three categories -- those affecting the cosmological clustering around the quasars, environmentally-dependent baryonic physics that are not accounted for in the simplistic modelling we use, and fainter average CO emission.

\subsection{Clustering of halos}
Because the stack luminosity is dominated by clustering effects, it is extremely sensitive to the number and mass (which translates directly to luminosity in our models) of the DM halos neighbouring those containing the quasars. In our modelling, this effect is treated through a combination of the QLF, onto which we abundance-match our simulated halos, and the `observational' cuts (in particular the cutoff luminosity $L'_\mathrm{cut}$) used to generate the actual mock catalogue being stacked upon. With our default parameter choices, we find roughly 250 DM halos fall into the stack aperture, in addition to the actual quasar host. The QLF is fairly well-characterized \citep[e.g.][]{shen2020_bolometricqlf, chaussidon2023_desiselection}, and statistics other than the QLF have also shown quasars to be highly clustered \citep{white2012_quasarclustering, timlin2018_quasarclustering}, so it is unlikely that our modelling is drastically incorrect. Still, small differences in the QLF can significantly affect the modelled stack luminosity (there is an average 18\% discrepancy between the two luminosity functions we test here). The model exclusion by our upper limit is tentative, and could vanish entirely with different choices of quasar modelling.

Modelling errors in our determination of the quasar velocity offsets could also change the number of halos falling into the stack aperture, by changing their systemic velocities. While we have attempted to account for the large bulk motions of gas into and out of quasars (\S\ref{sec:signal_attenuation:vel_uncert}), CO observations of quasars $z\sim 3$ are difficult to obtain, and our determination of the scatter in the relationship between the CO and optically-measured velocities is based on only 12 objects, measured using multiple CO lines. If the scatter is being underestimated, that would drive the modelled luminosities down and reduce the tension between the models and the data.

\subsection{Feedback and environmental effects}
The models we use to compare with the stack luminosity in this work are extremely simple. They assume a very basic galaxy-halo connection, taking only the DM mass of the halo (and possibly its redshift) into consideration when assigning luminosities for both the CO and the quasar tracers. Any other physics is ignored. In the context of a discrepancy between stack and model luminosities, any ignored physics which could result in galaxies in dense environments, or galaxies around quasars having decreased CO emission, could explain our results.  

Such environmental dependence is plausible -- galaxy properties such as SFR are well-known to correlate with their local density, potentially in a manner independent of their halo mass \citep[e.g.][]{park2007_sdssenvironment, peng2010_massandenvironment}. Traditionally, this environmental-dependence has been thought to matter most at small scales, but dependence on scales of $1 - 10\ \mathrm{Mpc}$ has been observed up to $z\sim 1$. \citep{park2007_sdssenvironment, wang2018_environmentquenching}. Quasars preferentially occupy dense environments; if environmental quenching is occurring in these environments, it would bring our modelled stack luminosities down. Additionally, quasar-specific feedback effects such as jets (which could extend up to several Mpc and have been observed at redshifts as high as $z\sim 5$; \citealt{oei2024_mpcjet, gloudemans2025_earlyjet}) could quench galaxies and disrupt CO emission around the eBOSS/DESI objects.

This sensitivity to large-scale environmental effects is an unique strength of the stack. While there are currently multiple degenerate explanations for the discrepancy between stack models and data, these degeneracies can be broken in the future with ancillary information. Once the CO autocorrelation power spectrum is measured, for example, this could be used to constrain the CO luminosity directly and any remaining discrepancy between models and data could be explained by environmental factors.

\subsection{Average CO luminosity}
The final explanation for the discrepancy between the data and models is that some of the models we present for cosmic CO are simply too bright. While considerable effort has been dedicated to realistically modelling the CO luminosity of even the faintest halos, only a very narrow mass range has actually been observed at $z\sim 3$ \citep[e.g.][]{riechers2019_coldz}, and the faint-end of the luminosity function has had to be largely inferred. The models we consider which assume only luminosity at the masses which have been observed, with little contribution from the faint end (Yang+2022, Chung+2022 flat+COLDz), are not in tension with our upper limit. The average cosmic CO luminosity could then be fainter without any disagreement with traditional surveys. Conversely, some LIM-based evidence has been presented favouring the Li+2016-Keating+2020 model \citep{keating2020_mmime, chung2024_COMAPS2_PaperIII}, but this evidence has primarily been tentative.

\section{Summary and Conclusions}
In this work, we have presented an update to the COMAP-eBOSS stack of \cite{dunne2024_ebossstacking}. In addition to the $3\times$ greater sensitivity in the stack that results from the increase in COMAP data since the ES data release, we have refined the stack methodology considerably, through a template-based matched filter. We place a $2\sigma$ upper limit on the CO emission surrounding 231 quasars from eBOSS and DESI of $\langle L'_\mathrm{CO} \rangle = 10.0\times 10^{10}\ \mathrm{K\ km\ s^{-1}\ pc^2}$. We compare this to models of the stack signal, varying both the prescription for CO emission and the QLF. We find that the COMAP-eBOSS/DESI upper limit excludes our brightest model of CO emission stacked on quasars (Li+2016-Keating+2020/Torralba-Torregrosa+2024) to $> 3\sigma$ even after accounting for experimental attenuation and variance from factors including sample variance, pipeline and experimental signal loss, velocity attenuation, model-dependence of the \meanlco\ extraction, and interloper emission in the LIM data. The COMAP fiducial/Torralba-Torregrosa+2024 model is also strongly disfavoured at $3.3\sigma$. 

However, a wide variety of disparate quasar and galaxy properties affect the stack in addition to the cosmic CO luminosity itself. This is because the stack is not a measure of the true average CO luminosity of the COMAP fields, instead probing only the CO luminosity of the regions surrounding quasars observable by eBOSS or DESI. While we rule out models for the stack emission, which combine CO and quasar modelling, we are thus not able to confidently rule out any model for cosmic CO emission from the stack alone. To truly isolate the cause of the observed discrepancy between data and models will require more detailed modelling and a much broader suite of LIM statistics to work from, including stacks on different galaxy tracers subject to independent environmental considerations, access to correlation data on more spatial scales from a cross-correlation power spectrum between the LIM data and the galaxy surveys, and ultimately a detection of cosmic CO in a full auto-correlation power spectrum as a comparison point. With these ancillary data points, the stack will be a powerful probe of large-scale environmental effects affecting galaxies.

% \section{Acknolwedgements}
\begin{acknowledgements}
This material is based upon work supported by the National Science Foundation under Grant Nos.\ 1517108, 1517288, 1517598, 1518282, 1910999, and 2206834, as well as by the Keck Institute for Space Studies under `The First Billion Years: A Technical Development Program for Spectral Line Observations'.

% COMAP authors
JG acknowledges support from the Keck Institute for Space Science, NSF AST-1517108 and University of Miami. GAH acknowledges the funding from the Dean’s Doctoral Scholarship by the University of Manchester. SM acknowledges support from the U.S.~Department of Defense through the Science, Mathematics, and Research for Transformation (SMART) Scholarship-for-Service Program.

    % SDSS/eBOSS/DESI
    This research made use of the SDSS-IV eBOSS survey. Funding for the Sloan Digital Sky Survey IV has been provided by the Alfred P. Sloan Foundation, the U.S. Department of Energy Office of Science, and the Participating Institutions. SDSS-IV acknowledges support and resources from the Center for High Performance Computing  at the University of Utah. The SDSS website is www.sdss4.org.
    
    SDSS-IV is managed by the Astrophysical Research Consortium for the Participating Institutions of the SDSS Collaboration including the Brazilian Participation Group, the Carnegie Institution for Science, Carnegie Mellon University, Center for Astrophysics | Harvard \& Smithsonian, the Chilean Participation Group, the French Participation Group, Instituto de Astrof\'isica de Canarias, The Johns Hopkins University, Kavli Institute for the Physics and Mathematics of the Universe (IPMU) / University of Tokyo, the Korean Participation Group, Lawrence Berkeley National Laboratory, Leibniz Institut f\"ur Astrophysik Potsdam (AIP),  Max-Planck-Institut f\"ur Astronomie (MPIA Heidelberg), Max-Planck-Institut f\"ur Astrophysik (MPA Garching), Max-Planck-Institut f\"ur Extraterrestrische Physik (MPE), National Astronomical Observatories of China, New Mexico State University, New York University, University of Notre Dame, Observat\'ario Nacional / MCTI, The Ohio State University, Pennsylvania State University, Shanghai Astronomical Observatory, United Kingdom Participation Group, Universidad Nacional Aut\'onoma de M\'exico, University of Arizona, University of Colorado Boulder, University of Oxford, University of Portsmouth, University of Utah, University of Virginia, University of Washington, University of Wisconsin, Vanderbilt University, and Yale University.

    This research used data obtained with the Dark Energy Spectroscopic Instrument (DESI). DESI construction and operations is managed by the Lawrence Berkeley National Laboratory. This material is based upon work supported by the U.S. Department of Energy, Office of Science, Office of High-Energy Physics, under Contract No. DE–AC02–05CH11231, and by the National Energy Research Scientific Computing Center, a DOE Office of Science User Facility under the same contract. Additional support for DESI was provided by the U.S. National Science Foundation (NSF), Division of Astronomical Sciences under Contract No. AST-0950945 to the NSF’s National Optical-Infrared Astronomy Research Laboratory; the Science and Technology Facilities Council of the United Kingdom; the Gordon and Betty Moore Foundation; the Heising-Simons Foundation; the French Alternative Energies and Atomic Energy Commission (CEA); the National Council of Humanities, Science and Technology of Mexico (CONAHCYT); the Ministry of Science and Innovation of Spain (MICINN), and by the DESI Member Institutions: www.desi.lbl.gov/collaborating-institutions. The DESI collaboration is honored to be permitted to conduct scientific research on I’oligam Du’ag (Kitt Peak), a mountain with particular significance to the Tohono O’odham Nation. Any opinions, findings, and conclusions or recommendations expressed in this material are those of the author(s) and do not necessarily reflect the views of the U.S. National Science Foundation, the U.S. Department of Energy, or any of the listed funding agencies.
\end{acknowledgements}

\bibliography{sample631}{}

\begin{thebibliography}{}
\expandafter\ifx\csname natexlab\endcsname\relax\def\natexlab#1{#1}\fi
\providecommand{\url}[1]{\href{#1}{#1}}
\providecommand{\dodoi}[1]{doi:~\href{http://doi.org/#1}{\nolinkurl{#1}}}
\providecommand{\doeprint}[1]{\href{http://ascl.net/#1}{\nolinkurl{http://ascl.net/#1}}}
\providecommand{\doarXiv}[1]{\href{https://arxiv.org/abs/#1}{\nolinkurl{https://arxiv.org/abs/#1}}}

\bibitem[{Ahumada {et~al.}(2020)Ahumada, Prieto, Almeida, Anders, Anderson, Andrews, Anguiano, Arcodia, Armengaud, Aubert, {et~al.}}]{eboss_dr16}
Ahumada, R., Prieto, C.~A., Almeida, A., {et~al.} 2020, \apjs, 249, 3

\bibitem[{{Aravena} {et~al.}(2019){Aravena}, {Decarli}, {G{\'o}nzalez-L{\'o}pez}, {Boogaard}, {Walter}, {Carilli}, {Popping}, {Weiss}, {Assef}, {Bacon}, {Bauer}, {Bertoldi}, {Bouwens}, {Contini}, {Cortes}, {Cox}, {da Cunha}, {Daddi}, {D{\'\i}az-Santos}, {Elbaz}, {Hodge}, {Inami}, {Ivison}, {Le F{\`e}vre}, {Magnelli}, {Oesch}, {Riechers}, {Smail}, {Somerville}, {Swinbank}, {Uzgil}, {van der Werf}, {Wagg}, \& {Wisotzki}}]{aravena2019_aspecs}
{Aravena}, M., {Decarli}, R., {G{\'o}nzalez-L{\'o}pez}, J., {et~al.} 2019, \apj, 882, 136, \dodoi{10.3847/1538-4357/ab30df}

\bibitem[{{Behroozi} {et~al.}(2019){Behroozi}, {Wechsler}, {Hearin}, \& {Conroy}}]{behroozi2019_universemachine}
{Behroozi}, P., {Wechsler}, R.~H., {Hearin}, A.~P., \& {Conroy}, C. 2019, \mnras, 488, 3143, \dodoi{10.1093/mnras/stz1182}

\bibitem[{{Bond} \& {Myers}(1996)}]{bondmeyers1996_peakpatchsims}
{Bond}, J.~R., \& {Myers}, S.~T. 1996, \apjs, 103, 1, \dodoi{10.1086/192267}

\bibitem[{Breysse \& Alexandroff(2019)}]{breysse2019_agnfeedback}
Breysse, P.~C., \& Alexandroff, R.~M. 2019, \mnras, 490, 260–273, \dodoi{10.1093/mnras/stz2534}

\bibitem[{Carilli \& Walter(2013)}]{carilliwalter2013_highzmoleculargas}
Carilli, C., \& Walter, F. 2013, Annual Review of Astronomy and Astrophysics, 51, 105, \dodoi{10.1146/annurev-astro-082812-140953}

\bibitem[{{CCAT-Prime Collaboration} {et~al.}(2023){CCAT-Prime Collaboration}, {Aravena}, {Austermann}, {Basu}, {Battaglia}, {Beringue}, {Bertoldi}, {Bigiel}, {Bond}, {Breysse}, {Broughton}, {Bustos}, {Chapman}, {Charmetant}, {Choi}, {Chung}, {Clark}, {Cothard}, {Crites}, {Dev}, {Douglas}, {Duell}, {D{\"u}nner}, {Ebina}, {Erler}, {Fich}, {Fissel}, {Foreman}, {Freundt}, {Gallardo}, {Gao}, {Garc{\'\i}a}, {Giovanelli}, {Golec}, {Groppi}, {Haynes}, {Henke}, {Hensley}, {Herter}, {Higgins}, {Hlo{\v{z}}ek}, {Huber}, {Huber}, {Hubmayr}, {Jackson}, {Johnstone}, {Karoumpis}, {Keating}, {Komatsu}, {Li}, {Magnelli}, {Matthews}, {Mauskopf}, {McMahon}, {Meerburg}, {Meyers}, {Muralidhara}, {Murray}, {Niemack}, {Nikola}, {Okada}, {Puddu}, {Riechers}, {Rosolowsky}, {Rossi}, {Rotermund}, {Roy}, {Sadavoy}, {Schaaf}, {Schilke}, {Scott}, {Simon}, {Sinclair}, {Sivakoff}, {Stacey}, {Stutz}, {Stutzki}, {Tahani}, {Thanjavur}, {Timmermann}, {Ullom}, {van Engelen}, {Vavagiakis}, {Vissers}, {Wheeler}, {White}, {Zhu}, \&
  {Zou}}]{ccatprime2023_overview}
{CCAT-Prime Collaboration}, {Aravena}, M., {Austermann}, J.~E., {et~al.} 2023, \apjs, 264, 7, \dodoi{10.3847/1538-4365/ac9838}

\bibitem[{{Chaussidon} {et~al.}(2023){Chaussidon}, {Y{\`e}che}, {Palanque-Delabrouille}, {Alexander}, {Yang}, {Ahlen}, {Bailey}, {Brooks}, {Cai}, {Chabanier}, {Davis}, {Dawson}, {de laMacorra}, {Dey}, {Dey}, {Eftekharzadeh}, {Eisenstein}, {Fanning}, {Font-Ribera}, {Gazta{\~n}aga}, {A Gontcho}, {Gonzalez-Morales}, {Guy}, {Herrera-Alcantar}, {Honscheid}, {Ishak}, {Jiang}, {Juneau}, {Kehoe}, {Kisner}, {Kov{\'a}cs}, {Kremin}, {Lan}, {Landriau}, {Le Guillou}, {Levi}, {Magneville}, {Martini}, {Meisner}, {Moustakas}, {Mu{\~n}oz-Guti{\'e}rrez}, {Myers}, {Newman}, {Nie}, {Percival}, {Poppett}, {Prada}, {Raichoor}, {Ravoux}, {Ross}, {Schlafly}, {Schlegel}, {Tan}, {Tarl{\'e}}, {Zhou}, {Zhou}, \& {Zou}}]{chaussidon2023_desiselection}
{Chaussidon}, E., {Y{\`e}che}, C., {Palanque-Delabrouille}, N., {et~al.} 2023, \apj, 944, 107, \dodoi{10.3847/1538-4357/acb3c2}

\bibitem[{{Chen} {et~al.}(2025){Chen}, {Cunnington}, {Pourtsidou}, {Wolz}, {Spinelli}, {Bernal}, {Barberi-Squarotti}, {Camera}, {Carucci}, {Fonseca}, {Grainge}, {Irfan}, {Santos}, \& {Wang}}]{chen2025_MeerKLASSstack}
{Chen}, Z., {Cunnington}, S., {Pourtsidou}, A., {et~al.} 2025, arXiv e-prints, arXiv:2504.03908, \dodoi{10.48550/arXiv.2504.03908}

\bibitem[{{Cheng} {et~al.}(2018){Cheng}, {Parsons}, {Kolopanis}, {Jacobs}, {Liu}, {Kohn}, {Aguirre}, {Pober}, {Ali}, {Bernardi}, {Bradley}, {Carilli}, {DeBoer}, {Dexter}, {Dillon}, {Klima}, {MacMahon}, {Moore}, {Nunhokee}, {Walbrugh}, \& {Walker}}]{cheng2018_paperbias}
{Cheng}, C., {Parsons}, A.~R., {Kolopanis}, M., {et~al.} 2018, \apj, 868, 26, \dodoi{10.3847/1538-4357/aae833}

\bibitem[{{Chowdhary} \& {Chatterjee}(2025)}]{chowdhary2025_quasarhod}
{Chowdhary}, A., \& {Chatterjee}, S. 2025, \apj, 992, 21, \dodoi{10.3847/1538-4357/adfb6d}

\bibitem[{{Chung} {et~al.}(2024{\natexlab{a}}){Chung}, {Chluba}, \& {Breysse}}]{chung2024_globalsignals}
{Chung}, D.~T., {Chluba}, J., \& {Breysse}, P.~C. 2024{\natexlab{a}}, \prd, 110, 023513, \dodoi{10.1103/PhysRevD.110.023513}

\bibitem[{{Chung} {et~al.}(2022){Chung}, {Breysse}, {Cleary}, {Ihle}, {Padmanabhan}, {Silva}, {Richard Bond}, {Borowska}, {Catha}, {Church}, {Dunne}, {Kristian Eriksen}, {Kristine Foss}, {Gaier}, {Ott Gundersen}, {Harper}, {Harris}, {Hensley}, {Hobbs}, {Keating}, {Kim}, {Lamb}, {Lawrence}, {Gahr Sturtzel Lunde}, {Murray}, {Pearson}, {Philip}, {Rasmussen}, {Readhead}, {Rennie}, {Stutzer}, {Uzgil}, {Viero}, {Watts}, {Wechsler}, {Kathrine Wehus}, {Woody}, \& {Comap Collaboration}}]{chung2021_comapforecasts}
{Chung}, D.~T., {Breysse}, P.~C., {Cleary}, K.~A., {et~al.} 2022, \apj, 933, 186, \dodoi{10.3847/1538-4357/ac63c7}

\bibitem[{{Chung} {et~al.}(2024{\natexlab{b}}){Chung}, {Breysse}, {Cleary}, {Dunne}, {Lunde}, {Padmanabhan}, {Stutzer}, {Tolgay}, {Bond}, {Church}, {Eriksen}, {Gaier}, {Gundersen}, {Harper}, {Harris}, {Hobbs}, {Ihle}, {Kim}, {Lamb}, {Lawrence}, {Murray}, {Pearson}, {Philip}, {Readhead}, {Rennie}, {Wehus}, {Woody}, \& {COMAP Collaboration}}]{chung2024_COMAPS2_PaperIII}
---. 2024{\natexlab{b}}, \aap, 691, A337, \dodoi{10.1051/0004-6361/202451122}

\bibitem[{{Cleary} {et~al.}(2022){Cleary}, {Borowska}, {Breysse}, {Catha}, {Chung}, {Church}, {Dickinson}, {Eriksen}, {Foss}, {Gundersen}, {Harper}, {Harris}, {Hobbs}, {Ihle}, {Kim}, {Kocz}, {Lamb}, {Lunde}, {Padmanabhan}, {Pearson}, {Philip}, {Powell}, {Rasmussen}, {Readhead}, {Rennie}, {Silva}, {Stutzer}, {Uzgil}, {Watts}, {Wehus}, {Woody}, {Basoalto}, {Bond}, {Dunne}, {Gaier}, {Hensley}, {Keating}, {Lawrence}, {Murray}, {Paladini}, {Reeves}, {Viero}, {Wechsler}, \& {Comap Collaboration}}]{cleary2021_comapoverview}
{Cleary}, K.~A., {Borowska}, J., {Breysse}, P.~C., {et~al.} 2022, \apj, 933, 182, \dodoi{10.3847/1538-4357/ac63cc}

\bibitem[{{CONCERTO Collaboration} {et~al.}(2020){CONCERTO Collaboration}, {Ade}, {Aravena}, {Barria}, {Beelen}, {Benoit}, {B{\'e}thermin}, {Bounmy}, {Bourrion}, {Bres}, {De Breuck}, {Calvo}, {Cao}, {Catalano}, {D{\'e}sert}, {Dur{\'a}n}, {Fasano}, {Fenouillet}, {Garcia}, {Garde}, {Goupy}, {Groppi}, {Hoarau}, {Lagache}, {Lambert}, {Leggeri}, {Levy-Bertrand}, {Mac{\'\i}as-P{\'e}rez}, {Mani}, {Marpaud}, {Mauskopf}, {Monfardini}, {Pisano}, {Ponthieu}, {Prieur}, {Roni}, {Roudier}, {Tourres}, \& {Tucker}}]{concerto2020_intro}
{CONCERTO Collaboration}, {Ade}, P., {Aravena}, M., {et~al.} 2020, \aap, 642, A60, \dodoi{10.1051/0004-6361/202038456}

\bibitem[{{Condon} {et~al.}(1998){Condon}, {Cotton}, {Greisen}, {Yin}, {Perley}, {Taylor}, \& {Broderick}}]{condon1998_nvss}
{Condon}, J.~J., {Cotton}, W.~D., {Greisen}, E.~W., {et~al.} 1998, \aj, 115, 1693, \dodoi{10.1086/300337}

\bibitem[{{Crites} {et~al.}(2014){Crites}, {Bock}, {Bradford}, {Chang}, {Cooray}, {Duband}, {Gong}, {Hailey-Dunsheath}, {Hunacek}, {Koch}, {Li}, {O'Brient}, {Prouve}, {Shirokoff}, {Silva}, {Staniszewski}, {Uzgil}, \& {Zemcov}}]{crites2014_timeSPIE}
{Crites}, A.~T., {Bock}, J.~J., {Bradford}, C.~M., {et~al.} 2014, in Society of Photo-Optical Instrumentation Engineers (SPIE) Conference Series, Vol. 9153, Millimeter, Submillimeter, and Far-Infrared Detectors and Instrumentation for Astronomy VII, ed. W.~S. {Holland} \& J.~{Zmuidzinas}, 91531W, \dodoi{10.1117/12.2057207}

\bibitem[{{Dawson} {et~al.}(2013){Dawson}, {Schlegel}, {Ahn}, {Anderson}, {Aubourg}, {Bailey}, {Barkhouser}, {Bautista}, {Beifiori}, {Berlind}, {Bhardwaj}, {Bizyaev}, {Blake}, {Blanton}, {Blomqvist}, {Bolton}, {Borde}, {Bovy}, {Brandt}, {Brewington}, {Brinkmann}, {Brown}, {Brownstein}, {Bundy}, {Busca}, {Carithers}, {Carnero}, {Carr}, {Chen}, {Comparat}, {Connolly}, {Cope}, {Croft}, {Cuesta}, {da Costa}, {Davenport}, {Delubac}, {de Putter}, {Dhital}, {Ealet}, {Ebelke}, {Eisenstein}, {Escoffier}, {Fan}, {Filiz Ak}, {Finley}, {Font-Ribera}, {G{\'e}nova-Santos}, {Gunn}, {Guo}, {Haggard}, {Hall}, {Hamilton}, {Harris}, {Harris}, {Ho}, {Hogg}, {Holder}, {Honscheid}, {Huehnerhoff}, {Jordan}, {Jordan}, {Kauffmann}, {Kazin}, {Kirkby}, {Klaene}, {Kneib}, {Le Goff}, {Lee}, {Long}, {Loomis}, {Lundgren}, {Lupton}, {Maia}, {Makler}, {Malanushenko}, {Malanushenko}, {Mandelbaum}, {Manera}, {Maraston}, {Margala}, {Masters}, {McBride}, {McDonald}, {McGreer}, {McMahon}, {Mena}, {Miralda-Escud{\'e}}, {Montero-Dorta},
  {Montesano}, {Muna}, {Myers}, {Naugle}, {Nichol}, {Noterdaeme}, {Nuza}, {Olmstead}, {Oravetz}, {Oravetz}, {Owen}, {Padmanabhan}, {Palanque-Delabrouille}, {Pan}, {Parejko}, {P{\^a}ris}, {Percival}, {P{\'e}rez-Fournon}, {P{\'e}rez-R{\`a}fols}, {Petitjean}, {Pfaffenberger}, {Pforr}, {Pieri}, {Prada}, {Price-Whelan}, {Raddick}, {Rebolo}, {Rich}, {Richards}, {Rockosi}, {Roe}, {Ross}, {Ross}, {Rossi}, {Rubi{\~n}o-Martin}, {Samushia}, {S{\'a}nchez}, {Sayres}, {Schmidt}, {Schneider}, {Sc{\'o}ccola}, {Seo}, {Shelden}, {Sheldon}, {Shen}, {Shu}, {Slosar}, {Smee}, {Snedden}, {Stauffer}, {Steele}, {Strauss}, {Streblyanska}, {Suzuki}, {Swanson}, {Tal}, {Tanaka}, {Thomas}, {Tinker}, {Tojeiro}, {Tremonti}, {Vargas Maga{\~n}a}, {Verde}, {Viel}, {Wake}, {Watson}, {Weaver}, {Weinberg}, {Weiner}, {West}, {White}, {Wood-Vasey}, {Yeche}, {Zehavi}, {Zhao}, \& {Zheng}}]{dawson2013_sdssbossreduction}
{Dawson}, K.~S., {Schlegel}, D.~J., {Ahn}, C.~P., {et~al.} 2013, \aj, 145, 10, \dodoi{10.1088/0004-6256/145/1/10}

\bibitem[{{Delubac} {et~al.}(2017){Delubac}, {Raichoor}, {Comparat}, {Jouvel}, {Kneib}, {Y{\`e}che}, {Zou}, {Brownstein}, {Abdalla}, {Dawson}, {Jullo}, {Myers}, {Newman}, {Percival}, {Prada}, {Ross}, {Schneider}, {Zhou}, {Zhou}, \& {Zhu}}]{delubac2017_ebosssystematics}
{Delubac}, T., {Raichoor}, A., {Comparat}, J., {et~al.} 2017, \mnras, 465, 1831, \dodoi{10.1093/mnras/stw2741}

\bibitem[{{DESI Collaboration} {et~al.}(2024){DESI Collaboration}, {Adame}, {Aguilar}, {Ahlen}, {Alam}, {Alexander}, {Alvarez}, {Alves}, {Anand}, {Andrade}, {Armengaud}, {Avila}, {Aviles}, {Awan}, {Bailey}, {Baltay}, {Bault}, {Behera}, {BenZvi}, {Beutler}, {Bianchi}, {Blake}, {Blum}, {Brieden}, {Brodzeller}, {Brooks}, {Brown}, {Buckley-Geer}, {Burtin}, {Calderon}, {Canning}, {Carnero Rosell}, {Cereskaite}, {Cervantes-Cota}, {Chabanier}, {Chaussidon}, {Chaves-Montero}, {Chen}, {Chen}, {Claybaugh}, {Cole}, {Cuceu}, {Davis}, {Dawson}, {de la Macorra}, {de Mattia}, {Deiosso}, {Demina}, {Dey}, {Dey}, {Ding}, {Doel}, {Edelstein}, {Eftekharzadeh}, {Eisenstein}, {Elliott}, {Fagrelius}, {Fanning}, {Ferraro}, {Ereza}, {Findlay}, {Flaugher}, {Font-Ribera}, {Forero-S{\'a}nchez}, {Forero-Romero}, {Frenk}, {Garcia-Quintero}, {Gazta{\~n}aga}, {Gil-Mar{\'\i}n}, {Gontcho}, {Gonzalez-Morales}, {Gonzalez-Perez}, {Gordon}, {Green}, {Gruen}, {Gsponer}, {Gutierrez}, {Guy}, {Hadzhiyska}, {Hahn}, {Hanif}, {Herrera-Alcantar},
  {Honscheid}, {Hou}, {Howlett}, {Huterer}, {Ir{\v{s}}i{\v{c}}}, {Ishak}, {Juneau}, {Kara{\c{c}}ayl{\i}}, {Kehoe}, {Kent}, {Kirkby}, {Kitaura}, {Kong}, {Kremin}, {Krolewski}, {Lai}, {Lan}, {Landriau}, {Lang}, {Lasker}, {Le Goff}, {Le Guillou}, {Leauthaud}, {Levi}, {Li}, {Lodha}, {Magneville}, {Manera}, {Margala}, {Martini}, {Maus}, {McDonald}, {Medina-Varela}, {Meisner}, {Mena-Fern{\'a}ndez}, {Miquel}, {Moon}, {Moore}, {Moustakas}, {Mudur}, {Mueller}, {Mu{\~n}oz-Guti{\'e}rrez}, {Myers}, {Nadathur}, {Napolitano}, {Neveux}, {Newman}, {Nguyen}, {Nie}, {Niz}, {Noriega}, {Padmanabhan}, {Paillas}, {Palanque-Delabrouille}, {Pan}, {Penmetsa}, {Percival}, {Pieri}, {Pinon}, {Poppett}, {Porredon}, {Prada}, {P{\'e}rez-Fern{\'a}ndez}, {P{\'e}rez-R{\`a}fols}, {Rabinowitz}, {Raichoor}, {Ram{\'\i}rez-P{\'e}rez}, {Ramirez-Solano}, {Rashkovetskyi}, {Ravoux}, {Rezaie}, {Rich}, {Rocher}, {Rockosi}, {Roe}, {Rosado-Marin}, {Ross}, {Rossi}, {Ruggeri}, {Ruhlmann-Kleider}, {Samushia}, {Sanchez}, {Saulder}, {Schlafly}, {Schlegel},
  {Scholte}, {Schubnell}, {Seo}, {Sharples}, {Silber}, {Slosar}, {Smith}, {Sprayberry}, {Tan}, {Tarl{\'e}}, {Trusov}, {Vaisakh}, {Valcin}, {Valdes}, {Vargas-Maga{\~n}a}, {Verde}, {Walther}, {Wang}, {Wang}, {Weaver}, {Weaverdyck}, {Wechsler}, {Weinberg}, {White}, {Wilson}, {Yu}, {Yu}, {Yuan}, {Y{\`e}che}, {Zaborowski}, {Zarrouk}, {Zhang}, \& {Zhao}}]{desisampledefs_2024}
{DESI Collaboration}, {Adame}, A.~G., {Aguilar}, J., {et~al.} 2024, arXiv e-prints, arXiv:2411.12020, \dodoi{10.48550/arXiv.2411.12020}

\bibitem[{{DESI Collaboration} {et~al.}(2025){DESI Collaboration}, {Abdul-Karim}, {Adame}, {Aguado}, {Aguilar}, {Ahlen}, {Alam}, {Aldering}, {Alexander}, {Alfarsy}, {Allen}, {Allende Prieto}, {Alves}, {Anand}, {Andrade}, {Armengaud}, {Avila}, {Aviles}, {Awan}, {Bailey}, {Baleato Lizancos}, {Ballester}, {Bault}, {Bautista}, {BenZvi}, {Beraldo e Silva}, {Bermejo-Climent}, {Beutler}, {Bianchi}, {Blake}, {Blum}, {Bolton}, {Bonici}, {Brieden}, {Brodzeller}, {Brooks}, {Buckley-Geer}, {Burtin}, {Canning}, {Carnero Rosell}, {Carr}, {Carrilho}, {Casas}, {Castander}, {Cereskaite}, {Cervantes-Cota}, {Chaussidon}, {Chaves-Montero}, {Chen}, {Chen}, {Claybaugh}, {Cole}, {Cooper}, {Cousinou}, {Cuceu}, {Davis}, {Dawson}, {de Belsunce}, {de la Cruz}, {de la Macorra}, {de Mattia}, {Deiosso}, {Della Costa}, {Demina}, {Demirbozan}, {DeRose}, {Dey}, {Dey}, {Ding}, {Ding}, {Doel}, {Douglass}, {Dowicz}, {Ebina}, {Edelstein}, {Eisenstein}, {Elbers}, {Emas}, {Escoffier}, {Fagrelius}, {Fan}, {Fanning}, {Fawcett},
  {Fern\'andez-Garc\'ia}, {Ferraro}, {Findlay}, {Font-Ribera}, {Forero-Romero}, {Forero-S\'anchez}, {Frenk}, {G\''ansicke}, {Galbany}, {Garc\'ia-Bellido}, {Garcia-Quintero}, {Garrison}, {Gazta\~{n}aga}, {Gil-Mar\textbackslash'in}, {Gnedin}, {Gontcho}, {Gonzalez-Morales}, {Gonzalez-Perez}, {Gordon}, {Graur}, {Green}, {Gruen}, {Gsponer}, {Guandalin}, {Gutierrez}, {Guy}, {Hahn}, {Han}, {Han}, {He}, {Herrera-Alcantar}, {Honscheid}, {Hou}, {Howlett}, {Huterer}, {Ir{\v{s}}i{\v{c}}}, {Ishak}, {Jacques}, {Jimenez}, {Jing}, {Joachimi}, {Joudaki}, {Joyce}, {Jullo}, {Juneau}, {Kara{\c{c}}ayl{\i}}, {Karim}, {Kehoe}, {Kent}, {Khederlarian}, {Kirkby}, {Kisner}, {Kitaura}, {Kizhuprakkat}, {Kong}, {Koposov}, {Kremin}, {Krolewski}, {Lahav}, {Lai}, {Lamman}, {Lan}, {Landriau}, {Lang}, {Lange}, {Lasker}, {Le Goff}, {Le Guillou}, {Leauthaud}, {Levi}, {Li}, {Li}, {Lodha}, {Lokken}, {Luo}, {Magneville}, {Manera}, {Manser}, {Margala}, {Martini}, {Maus}, {McCullough}, {McDonald}, {Medina}, {Medina-Varela}, {Meisner},
  {Mena-Fern\'andez}, {Menegas}, {Mezcua}, {Miquel}, {Montero-Camacho}, {Moon}, {Moustakas}, {Mu\~{n}oz-Gut\'errez}, {Mu\~{n}oz-Santos}, {Myers}, {Myles}, {Nadathur}, {Najita}, {Napolitano}, {Newman}, {Nikakhtar}, {Nikutta}, {Niz}, {Noriega}, {Padmanabhan}, {Paillas}, {Palanque-Delabrouille}, {Palmese}, {Pan}, {Pan}, {Parkinson}, {Peacock}, {Percival}, {P\'erez-Fern\'andez}, {P\'erez-R\`afols}, \& {Peterson}}]{desidr1_2025}
{DESI Collaboration}, {Abdul-Karim}, M., {Adame}, A.~G., {et~al.} 2025, arXiv e-prints, arXiv:2503.14745, \dodoi{10.48550/arXiv.2503.14745}

\bibitem[{{Dor{\'e}} {et~al.}(2014){Dor{\'e}}, {Bock}, {Ashby}, {Capak}, {Cooray}, {de Putter}, {Eifler}, {Flagey}, {Gong}, {Habib}, {Heitmann}, {Hirata}, {Jeong}, {Katti}, {Korngut}, {Krause}, {Lee}, {Masters}, {Mauskopf}, {Melnick}, {Mennesson}, {Nguyen}, {{\"O}berg}, {Pullen}, {Raccanelli}, {Smith}, {Song}, {Tolls}, {Unwin}, {Venumadhav}, {Viero}, {Werner}, \& {Zemcov}}]{dore2014_spherexintro}
{Dor{\'e}}, O., {Bock}, J., {Ashby}, M., {et~al.} 2014, arXiv e-prints, arXiv:1412.4872, \dodoi{10.48550/arXiv.1412.4872}

\bibitem[{{Dunne} {et~al.}(2024){Dunne}, {Cleary}, {Breysse}, {Chung}, {Ihle}, {Bond}, {Eriksen}, {Gundersen}, {Keating}, {Kim}, {Lunde}, {Murray}, {Padmanabhan}, {Philip}, {Stutzer}, {Tolgay}, {Wehus}, {Church}, {Gaier}, {Harris}, {Hobbs}, {Lamb}, {Lawrence}, {Readhead}, \& {Woody}}]{dunne2024_ebossstacking}
{Dunne}, D.~A., {Cleary}, K.~A., {Breysse}, P.~C., {et~al.} 2024, \apj, 965, 7, \dodoi{10.3847/1538-4357/ad2dfc}

\bibitem[{{Dunne} {et~al.}(2025){Dunne}, {Cleary}, {Breysse}, {Chung}, {Ihle}, {Lunde}, {Padmanabhan}, {Stutzer}, {Bond}, {Gundersen}, {Kim}, \& {Readhead}}]{dunne2025_stacktheory}
---. 2025, arXiv e-prints, arXiv:2503.21743, \dodoi{10.48550/arXiv.2503.21743}

\bibitem[{{Gloudemans} {et~al.}(2025){Gloudemans}, {Sweijen}, {Morabito}, {Farina}, {Duncan}, {Harikane}, {R{\"o}ttgering}, {Saxena}, \& {Schindler}}]{gloudemans2025_earlyjet}
{Gloudemans}, A.~J., {Sweijen}, F., {Morabito}, L.~K., {et~al.} 2025, \apjl, 980, L8, \dodoi{10.3847/2041-8213/ad9609}

\bibitem[{{Harris} {et~al.}(2016){Harris}, {Jensen}, {Suzuki}, {Bautista}, {Dawson}, {Vivek}, {Brownstein}, {Ge}, {Hamann}, {Herbst}, {Jiang}, {Moran}, {Myers}, {Olmstead}, \& {Schneider}}]{harris2016_quasarcompositespectrum}
{Harris}, D.~W., {Jensen}, T.~W., {Suzuki}, N., {et~al.} 2016, \aj, 151, 155, \dodoi{10.3847/0004-6256/151/6/155}

\bibitem[{{Hill} {et~al.}(2021){Hill}, {Lee}, {MacQueen}, {Kelz}, {Drory}, {Vattiat}, {Good}, {Ramsey}, {Kriel}, {Peterson}, {DePoy}, {Gebhardt}, {Marshall}, {Tuttle}, {Bauer}, {Chonis}, {Fabricius}, {Froning}, {H{\"a}user}, {Indahl}, {Jahn}, {Landriau}, {Leck}, {Montesano}, {Prochaska}, {Snigula}, {Zeimann}, {Bryant}, {Damm}, {Fowler}, {Janowiecki}, {Martin}, {Mrozinski}, {Odewahn}, {Rostopchin}, {Shetrone}, {Spencer}, {Mentuch Cooper}, {Armandroff}, {Bender}, {Dalton}, {Hopp}, {Komatsu}, {Nicklas}, {Ramsey}, {Roth}, {Schneider}, {Sneden}, \& {Steinmetz}}]{hill2021_hetdexvirus}
{Hill}, G.~J., {Lee}, H., {MacQueen}, P.~J., {et~al.} 2021, \aj, 162, 298, \dodoi{10.3847/1538-3881/ac2c02}

\bibitem[{{Hill} {et~al.}(2024){Hill}, {Lee}, {Good}, {Drory}, {Vestuto}, {Vattiat}, {Kyte}, {Zeimann}, {Smith}, {Indahl}, \& {Haubitz-Reinke}}]{hill2024_hetdexvirus2}
{Hill}, G.~J., {Lee}, H., {Good}, J.~M., {et~al.} 2024, in Society of Photo-Optical Instrumentation Engineers (SPIE) Conference Series, Vol. 13096, Ground-based and Airborne Instrumentation for Astronomy X, ed. J.~J. {Bryant}, K.~{Motohara}, \& J.~R.~D. {Vernet}, 130960F, \dodoi{10.1117/12.3017716}

\bibitem[{{Hill} {et~al.}(2019){Hill}, {Chapman}, {Scott}, {Smail}, {Steidel}, {Krips}, {Babul}, {Berg}, {Bertoldi}, {Gao}, {Lacaille}, {Matsuda}, {Ross}, {Rudie}, \& {Trainor}}]{hill2019}
{Hill}, R., {Chapman}, S.~C., {Scott}, D., {et~al.} 2019, MNRAS, 485, 753, \dodoi{10.1093/mnras/stz429}

\bibitem[{Hinshaw {et~al.}(2013)Hinshaw, Larson, Komatsu, Spergel, Bennett, Dunkley, Nolta, Halpern, Hill, Odegard, Page, Smith, Weiland, Gold, Jarosik, Kogut, Limon, Meyer, Tucker, Wollack, \& Wright}]{Hinshaw_2013}
Hinshaw, G., Larson, D., Komatsu, E., {et~al.} 2013, \apjs, 208, 19, \dodoi{10.1088/0067-0049/208/2/19}

\bibitem[{{Hopkins} {et~al.}(2006){Hopkins}, {Hernquist}, {Cox}, {Di Matteo}, {Robertson}, \& {Springel}}]{hopkins2006_agnmodel}
{Hopkins}, P.~F., {Hernquist}, L., {Cox}, T.~J., {et~al.} 2006, \apjs, 163, 1, \dodoi{10.1086/499298}

\bibitem[{{Hopkins} {et~al.}(2008){Hopkins}, {Hernquist}, {Cox}, \& {Kere{\v{s}}}}]{hopkins2008_mergersandquasars}
{Hopkins}, P.~F., {Hernquist}, L., {Cox}, T.~J., \& {Kere{\v{s}}}, D. 2008, \apjs, 175, 356, \dodoi{10.1086/524362}

\bibitem[{{Kamenetzky} {et~al.}(2016){Kamenetzky}, {Rangwala}, {Glenn}, {Maloney}, \& {Conley}}]{kamenetzky2016_cofir}
{Kamenetzky}, J., {Rangwala}, N., {Glenn}, J., {Maloney}, P.~R., \& {Conley}, A. 2016, \apj, 829, 93, \dodoi{10.3847/0004-637X/829/2/93}

\bibitem[{{Keating} {et~al.}(2020){Keating}, {Marrone}, {Bower}, \& {Keenan}}]{keating2020_mmime}
{Keating}, G.~K., {Marrone}, D.~P., {Bower}, G.~C., \& {Keenan}, R.~P. 2020, \apj, 901, 141, \dodoi{10.3847/1538-4357/abb08e}

\bibitem[{Keating {et~al.}(2015)Keating, Bower, Marrone, DeBoer, Heiles, Chang, Carlstrom, Greer, Hawkins, Lamb, {et~al.}}]{keating2015_copss1}
Keating, G.~K., Bower, G.~C., Marrone, D.~P., {et~al.} 2015, \apj, 814, 140

\bibitem[{{Keenan} {et~al.}(2022){Keenan}, {Keating}, \& {Marrone}}]{keenan2021_copssstack}
{Keenan}, R.~P., {Keating}, G.~K., \& {Marrone}, D.~P. 2022, \apj, 927, 161, \dodoi{10.3847/1538-4357/ac4888}

\bibitem[{{Lamb} {et~al.}(2022){Lamb}, {Cleary}, {Woody}, {Catha}, {Chung}, {Gundersen}, {Harper}, {Harris}, {Hobbs}, {Ihle}, {Kocz}, {Pearson}, {Philip}, {Powell}, {Basoalto}, {Bond}, {Borowska}, {Breysse}, {Church}, {Dickinson}, {Dunne}, {Eriksen}, {Foss}, {Gaier}, {Kim}, {Lawrence}, {Lunde}, {Padmanabhan}, {Rasmussen}, {Readhead}, {Reeves}, {Rennie}, {Stutzer}, {Viero}, {Watts}, {Wehus}, \& {Comap Collaboration}}]{lamb2021_instrument}
{Lamb}, J.~W., {Cleary}, K.~A., {Woody}, D.~P., {et~al.} 2022, \apj, 933, 183, \dodoi{10.3847/1538-4357/ac63c6}

\bibitem[{{Lenki{\'c}} {et~al.}(2020){Lenki{\'c}}, {Bolatto}, {F{\"o}rster Schreiber}, {Tacconi}, {Neri}, {Combes}, {Walter}, {Garc{\'\i}a-Burillo}, {Genzel}, {Lutz}, \& {Cooper}}]{lenkic2020_phibss2}
{Lenki{\'c}}, L., {Bolatto}, A.~D., {F{\"o}rster Schreiber}, N.~M., {et~al.} 2020, \aj, 159, 190, \dodoi{10.3847/1538-3881/ab7458}

\bibitem[{{Li} {et~al.}(2016){Li}, {Wechsler}, {Devaraj}, \& {Church}}]{li2016_comodelling}
{Li}, T.~Y., {Wechsler}, R.~H., {Devaraj}, K., \& {Church}, S.~E. 2016, \apj, 817, 169, \dodoi{10.3847/0004-637X/817/2/169}

\bibitem[{{Lunde} {et~al.}(2024){Lunde}, {Stutzer}, {Breysse}, {Chung}, {Cleary}, {Dunne}, {Eriksen}, {Harper}, {Ihle}, {Lamb}, {Pearson}, {Philip}, {Wehus}, {Woody}, {Bond}, {Church}, {Gaier}, {Gundersen}, {Harris}, {Hobbs}, {Kim}, {Lawrence}, {Murray}, {Padmanabhan}, {Readhead}, {Rennie}, {Tolgay}, \& {COMAP Collaboration}}]{lunde2024_COMAPS2_PaperI}
{Lunde}, J.~G.~S., {Stutzer}, N.~O., {Breysse}, P.~C., {et~al.} 2024, \aap, 691, A335, \dodoi{10.1051/0004-6361/202451121}

\bibitem[{{Lunde} {et~al.}(in prep.){Lunde}, {Cleary}, {Dickinson}, {Dunne}, {Harper}, {Hoerning}, {Ihle}, {Lamb}, {Rennie}, \& {Stutzer}}]{lunde2025_gaincal}
{Lunde}, J.~G.~S., {Cleary}, K.~A., {Dickinson}, C., {et~al.} in prep., \aap

\bibitem[{Lyke {et~al.}(2020)Lyke, Higley, McLane, Schurhammer, Myers, Ross, Dawson, Chabanier, Martini, \& Busca}]{sdssdr16}
Lyke, B.~W., Higley, A.~N., McLane, J.~N., {et~al.} 2020, \apjs, 250, 8, \dodoi{10.3847/1538-4365/aba623}

\bibitem[{{Madau} \& {Dickinson}(2014)}]{madaudickinson2014_sfhistory}
{Madau}, P., \& {Dickinson}, M. 2014, \araa, 52, 415, \dodoi{10.1146/annurev-astro-081811-125615}

\bibitem[{{Mansfield} {et~al.}(in prep.){Mansfield}, {Chung}, \& {Dunne}}]{mansfield2025_adaptivephot}
{Mansfield}, E.~M., {Chung}, D.~T., \& {Dunne}, D.~A. in prep., \aap

\bibitem[{{Mu{\~n}oz-Elgueta} {et~al.}(2022){Mu{\~n}oz-Elgueta}, {Arrigoni Battaia}, {Kauffmann}, {De Breuck}, {Garc{\'\i}a-Vergara}, {Zanella}, {Farina}, \& {Decarli}}]{munozelgueta2022_apexqsos}
{Mu{\~n}oz-Elgueta}, N., {Arrigoni Battaia}, F., {Kauffmann}, G., {et~al.} 2022, \mnras, 511, 1462, \dodoi{10.1093/mnras/stac041}

\bibitem[{{Oei} {et~al.}(2024){Oei}, {Hardcastle}, {Timmerman}, {Gast}, {Botteon}, {Rodriguez}, {Stern}, {Calistro Rivera}, {van Weeren}, {R{\"o}ttgering}, {Intema}, {de Gasperin}, \& {Djorgovski}}]{oei2024_mpcjet}
{Oei}, M. S.~S.~L., {Hardcastle}, M.~J., {Timmerman}, R., {et~al.} 2024, \nat, 633, 537, \dodoi{10.1038/s41586-024-07879-y}

\bibitem[{{Padmanabhan}(2018)}]{padmanabhan2018_comodel}
{Padmanabhan}, H. 2018, \mnras, 475, 1477, \dodoi{10.1093/mnras/stx3250}

\bibitem[{{Palanque-Delabrouille} {et~al.}(2016){Palanque-Delabrouille}, {Magneville}, {Y{\`e}che}, {P{\^a}ris}, {Petitjean}, {Burtin}, {Dawson}, {McGreer}, {Myers}, {Rossi}, {Schlegel}, {Schneider}, {Streblyanska}, \& {Tinker}}]{palanquedelabrouille2016_ebosslumfunc}
{Palanque-Delabrouille}, N., {Magneville}, C., {Y{\`e}che}, C., {et~al.} 2016, \aap, 587, A41, \dodoi{10.1051/0004-6361/201527392}

\bibitem[{{Park} {et~al.}(2007){Park}, {Choi}, {Vogeley}, {Gott}, {Blanton}, \& {SDSS Collaboration}}]{park2007_sdssenvironment}
{Park}, C., {Choi}, Y.-Y., {Vogeley}, M.~S., {et~al.} 2007, \apj, 658, 898, \dodoi{10.1086/511059}

\bibitem[{{Peng} {et~al.}(2010){Peng}, {Lilly}, {Kova{\v{c}}}, {Bolzonella}, {Pozzetti}, {Renzini}, {Zamorani}, {Ilbert}, {Knobel}, {Iovino}, {Maier}, {Cucciati}, {Tasca}, {Carollo}, {Silverman}, {Kampczyk}, {de Ravel}, {Sanders}, {Scoville}, {Contini}, {Mainieri}, {Scodeggio}, {Kneib}, {Le F{\`e}vre}, {Bardelli}, {Bongiorno}, {Caputi}, {Coppa}, {de la Torre}, {Franzetti}, {Garilli}, {Lamareille}, {Le Borgne}, {Le Brun}, {Mignoli}, {Perez Montero}, {Pello}, {Ricciardelli}, {Tanaka}, {Tresse}, {Vergani}, {Welikala}, {Zucca}, {Oesch}, {Abbas}, {Barnes}, {Bordoloi}, {Bottini}, {Cappi}, {Cassata}, {Cimatti}, {Fumana}, {Hasinger}, {Koekemoer}, {Leauthaud}, {Maccagni}, {Marinoni}, {McCracken}, {Memeo}, {Meneux}, {Nair}, {Porciani}, {Presotto}, \& {Scaramella}}]{peng2010_massandenvironment}
{Peng}, Y.-j., {Lilly}, S.~J., {Kova{\v{c}}}, K., {et~al.} 2010, \apj, 721, 193, \dodoi{10.1088/0004-637X/721/1/193}

\bibitem[{{Richardson} {et~al.}(2012){Richardson}, {Zheng}, {Chatterjee}, {Nagai}, \& {Shen}}]{richardson2012_quasarHOD}
{Richardson}, J., {Zheng}, Z., {Chatterjee}, S., {Nagai}, D., \& {Shen}, Y. 2012, \apj, 755, 30, \dodoi{10.1088/0004-637X/755/1/30}

\bibitem[{{Riechers} {et~al.}(2019){Riechers}, {Pavesi}, {Sharon}, {Hodge}, {Decarli}, {Walter}, {Carilli}, {Aravena}, {da Cunha}, {Daddi}, {Dickinson}, {Smail}, {Capak}, {Ivison}, {Sargent}, {Scoville}, \& {Wagg}}]{riechers2019_coldz}
{Riechers}, D.~A., {Pavesi}, R., {Sharon}, C.~E., {et~al.} 2019, \apj, 872, 7, \dodoi{10.3847/1538-4357/aafc27}

\bibitem[{{Sanders} {et~al.}(1988){Sanders}, {Soifer}, {Elias}, {Madore}, {Matthews}, {Neugebauer}, \& {Scoville}}]{sanders1988_quasarorigins}
{Sanders}, D.~B., {Soifer}, B.~T., {Elias}, J.~H., {et~al.} 1988, \apj, 325, 74, \dodoi{10.1086/165983}

\bibitem[{{Schechter}(1976)}]{schechter1976}
{Schechter}, P. 1976, \apj, 203, 297, \dodoi{10.1086/154079}

\bibitem[{{Shen} {et~al.}(2020){Shen}, {Hopkins}, {Faucher-Gigu{\`e}re}, {Alexander}, {Richards}, {Ross}, \& {Hickox}}]{shen2020_bolometricqlf}
{Shen}, X., {Hopkins}, P.~F., {Faucher-Gigu{\`e}re}, C.-A., {et~al.} 2020, \mnras, 495, 3252, \dodoi{10.1093/mnras/staa1381}

\bibitem[{Solomon {et~al.}(1997)Solomon, Downes, Radford, \& Barrett}]{Solomon_1997}
Solomon, P.~M., Downes, D., Radford, S. J.~E., \& Barrett, J.~W. 1997, \apj, 478, 144, \dodoi{10.1086/303765}

\bibitem[{{Somerville} {et~al.}(2015){Somerville}, {Popping}, \& {Trager}}]{somerville2015_santacruz}
{Somerville}, R.~S., {Popping}, G., \& {Trager}, S.~C. 2015, \mnras, 453, 4337, \dodoi{10.1093/mnras/stv1877}

\bibitem[{{Stein} {et~al.}(2019){Stein}, {Alvarez}, \& {Bond}}]{stein2019_peakpatchsims}
{Stein}, G., {Alvarez}, M.~A., \& {Bond}, J.~R. 2019, \mnras, 483, 2236, \dodoi{10.1093/mnras/sty3226}

\bibitem[{{Stutzer} {et~al.}(2024){Stutzer}, {Lunde}, {Breysse}, {Chung}, {Cleary}, {Dunne}, {Eriksen}, {Ihle}, {Padmanabhan}, {Tolgay}, {Wehus}, {Bond}, {Church}, {Gaier}, {Gundersen}, {Harris}, {Harper}, {Hobbs}, {Kim}, {Lamb}, {Lawrence}, {Murray}, {Pearson}, {Philip}, {Readhead}, {Rennie}, \& {Woody}}]{stutzer2024_COMAPS2_PaperII}
{Stutzer}, N.~O., {Lunde}, J.~G.~S., {Breysse}, P.~C., {et~al.} 2024, \aap, 691, A336, \dodoi{10.1051/0004-6361/202451123}

\bibitem[{{Timlin} {et~al.}(2018){Timlin}, {Ross}, {Richards}, {Myers}, {Pellegrino}, {Bauer}, {Lacy}, {Schneider}, {Wollack}, \& {Zakamska}}]{timlin2018_quasarclustering}
{Timlin}, J.~D., {Ross}, N.~P., {Richards}, G.~T., {et~al.} 2018, \apj, 859, 20, \dodoi{10.3847/1538-4357/aab9ac}

\bibitem[{{Torralba-Torregrosa} {et~al.}(2024){Torralba-Torregrosa}, {Renard}, {Spinoso}, {Arnalte-Mur}, {Gurung-L{\'o}pez}, {Fern{\'a}ndez-Soto}, {Gazta{\~n}aga}, {Navarro-Giron{\'e}s}, {Cai}, {Carretero}, {Castander}, {Eriksen}, {Garcia-Bellido}, {Hildebrandt}, {Hoekstra}, {Miquel}, {Sanchez}, {Tallada-Cresp{\'\i}}, {De Vicente}, \& {Fernandez}}]{torralba-torregrosa2024_quasarlyalumfunc}
{Torralba-Torregrosa}, A., {Renard}, P., {Spinoso}, D., {et~al.} 2024, \aap, 690, A388, \dodoi{10.1051/0004-6361/202451675}

\bibitem[{{Vieira} {et~al.}(2020){Vieira}, {Aguirre}, {Bradford}, {Filippini}, {Groppi}, {Marrone}, {Bethermin}, {Chang}, {Devlin}, {Dore}, {Fu}, {Hailey Dunsheath}, {Holder}, {Keating}, {Keenan}, {Kovetz}, {Lagache}, {Mauskopf}, {Narayanan}, {Popping}, {Shirokoff}, {Somerville}, {Trumper}, {Uzgil}, \& {Zmuidzinas}}]{viera2020_tim}
{Vieira}, J., {Aguirre}, J., {Bradford}, C.~M., {et~al.} 2020, arXiv e-prints, arXiv:2009.14340, \dodoi{10.48550/arXiv.2009.14340}

\bibitem[{{Wang} {et~al.}(2018){Wang}, {Mo}, {Chen}, {Yang}, {Yang}, {Wang}, {van den Bosch}, {Jing}, {Kang}, {Lin}, {Lim}, {Huang}, {Lu}, {Li}, {Cui}, {Zhang}, {Tweed}, {Wei}, {Li}, \& {Shi}}]{wang2018_environmentquenching}
{Wang}, H., {Mo}, H.~J., {Chen}, S., {et~al.} 2018, \apj, 852, 31, \dodoi{10.3847/1538-4357/aa9e01}

\bibitem[{{White} {et~al.}(2012){White}, {Myers}, {Ross}, {Schlegel}, {Hennawi}, {Shen}, {McGreer}, {Strauss}, {Bolton}, {Bovy}, {Fan}, {Miralda-Escude}, {Palanque-Delabrouille}, {Paris}, {Petitjean}, {Schneider}, {Viel}, {Weinberg}, {Yeche}, {Zehavi}, {Pan}, {Snedden}, {Bizyaev}, {Brewington}, {Brinkmann}, {Malanushenko}, {Malanushenko}, {Oravetz}, {Simmons}, {Sheldon}, \& {Weaver}}]{white2012_quasarclustering}
{White}, M., {Myers}, A.~D., {Ross}, N.~P., {et~al.} 2012, \mnras, 424, 933, \dodoi{10.1111/j.1365-2966.2012.21251.x}

\bibitem[{{Yang} {et~al.}(2022){Yang}, {Popping}, {Somerville}, {Pullen}, {Breysse}, \& {Maniyar}}]{yang2022_SAMs}
{Yang}, S., {Popping}, G., {Somerville}, R.~S., {et~al.} 2022, \apj, 929, 140, \dodoi{10.3847/1538-4357/ac5d57}

\end{thebibliography}
\bibliographystyle{aasjournal}

\begin{appendix}

\section{Signal injection boost strength}\label{app:signal_injection_boost}
We determine pipeline signal loss and extract the stack \meanlco\ using injection simulations, where we inject simulated CO LIM cubes into subsets of the full COMAP data, boosting the injected signal as we do. As discussed in \S\ref{sec:methodology:injections:templates} and \S\ref{sec:methodology:extraction}, we use a boost strength of 50. This yields high S/N ratios in the injected signal when stacked and averaged over 100 iterations (S/N ratios of 49-67 in the central voxel, depending on field). Here, we justify that this boost is also small enough not to cause non-linearities in any of the pipeline steps, or to interact with existing signal in the data.

Ensuring that the boosted injected signal will not affect the TOD pipeline is relatively simple -- we find that the brightest voxel of the simulated emission is still roughly four orders of magnitude below the system temperature even after boosting by a factor of 50. The simulated TOD signal will thus be well below the noise, and the TOD pipeline will not be affected by the injection of this signal. The other potentially non-linear pipeline step is the PCA filter applied after map-making \citep[][]{lunde2024_COMAPS2_PaperI}. The map S/N ratio is greatly increased compared to the S/N ratio of the TOD, so it is possible for the signal to be in the non-linear regime of this filter even when it is well below the noise in the TOD. \cite{lunde2024_COMAPS2_PaperI} show how the filter behaves with the voxel-level S/N of the maps, $\langle \mathrm{SNR_{vox}} \rangle$. They find that the map-level PCA filter behaves linearly with respect to the CO signal up to $\langle \mathrm{SNR_{vox}} \rangle = 0.02$. The boosted, signal-injected data subsets we use in this work have $\langle \mathrm{SNR_{vox}} \rangle \sim 0.005$, so we are well within this linear regime. We also note that our best estimates for the real signal will be well within the linear regime of this filter. The unboosted fiducial CO model would be at a $\langle \mathrm{SNR_{vox}} \rangle = 2\times10^{-4}$ in each subset of the full COMAP S2 dataset, and this rises to $\langle \mathrm{SNR_{vox}} \rangle = 0.002$ after averaging the 100 subsets together. We therefore do not expect either the signal loss analysis or the stack itself to be affected by pipeline nonlinearities. As an additional check that there is no signal being removed by the map-level PCA filter, we stack on the PCA modes removed by the filter. We find no signal.

We also note that signal-injection simulations have been shown to give inaccurate results when there are interactions between real and injected signal \citep[][]{cheng2018_paperbias}. This is not an issue when the injected signal is considerably stronger than the real signal in the data. While we do not know the actual strength of the signal in the COMAP data, this work suggests that it is likely weaker than the Li+2016-Keating+2020 model, and may be weaker than the COMAP fiducial model. Assuming the fiducial model is correct, the injected signal will be on average $50\times$ brighter than the real signal in the data cubes. Stacking introduces an additional prior on the position of the signal, so even if the two were of equal brightness the real signal would be at least $9.5\times$ fainter than the injected signal in a stack on the injection catalogue. This means that overall, the real signal must be at least $500\times$ fainter than the injected, so we do not expect biasing due to the injection-based simulation strategy.

\section{Pointing verification using continuum radio sources}\label{app:verification}
\begin{figure*}
    \centering
    \includegraphics[width=0.95\linewidth]{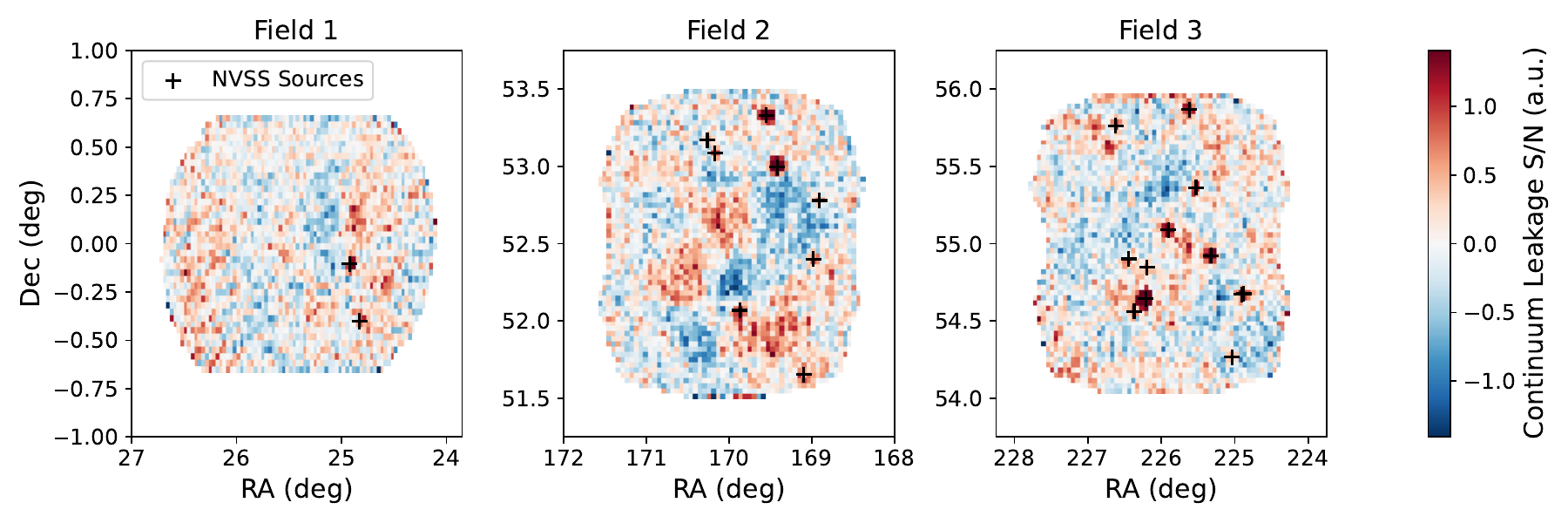}
    \caption{Continuum-leakage maps of the three COMAP fields, with radio continuum sources from NVSS \citep[][]{condon1998_nvss} overplotted.}
    \label{fig:continuum_sources}
\end{figure*}

We verify the COMAP pointing and the stack coordinate system by performing a stack on known radio continuum sources from the NRAO VLA Sky Survey \citep[][NVSS]{condon1998_nvss}. The COMAP pipeline is designed to reject continuum, and does so to a high degree of significance \citep[][]{lunde2024_COMAPS2_PaperI}. However, continuum-leakage maps can be generated by purposefully accounting for the system temperature spikes incorrectly during time-domain processing \citep{lunde2025_gaincal}.  These maps can then be averaged across the frequency axis to give continuum. While the flux density in these maps are extremely difficult to calibrate, the spatial distribution of any emission present is accurate. Visible in these maps are clear signatures of the Cosmic Microwave Background, as well as several bright radio continuum sources (Figure \ref{fig:continuum_sources}).

\begin{figure}
    \centering
    \includegraphics[width=0.8\linewidth]{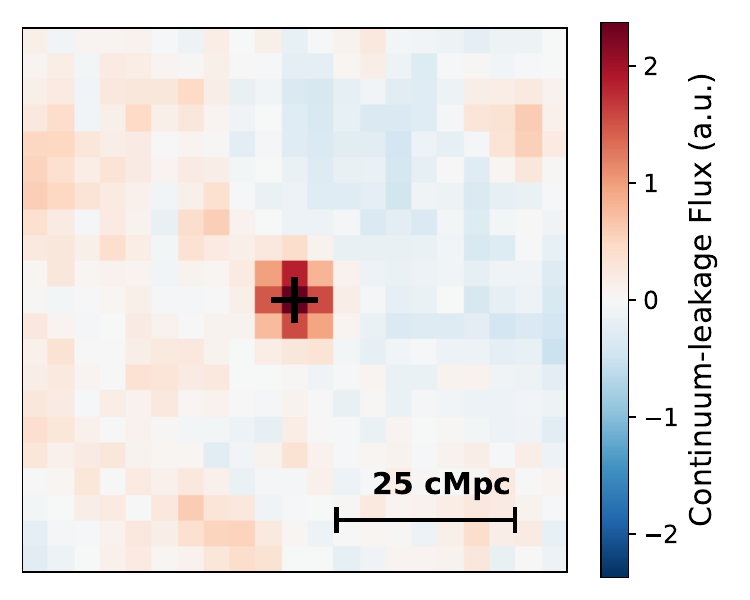}
    \caption{Stack of 22 NVSS radio sources on the COMAP continuum-leakage maps. The resulting image is roughly beam-sized, indicating good pointing accuracy. The scale bar is shown for $z=2.62$ (the median redshift of the eBOSS/DESI sources). It corresponds to $\sim14'$.}
    \label{fig:continuum_stack}
\end{figure}

There are 22 sources from NVSS that fall within the COMAP fields. We perform a stack on these 22 sources, using the frequency-averaged map. The resulting angular profile is shown in Figure \ref{fig:continuum_stack}. We find a source profile that is somewhat more spread-out than the $4.5'$ Gaussian to be expected from stacking on sources exactly centred in their spaxels and broadened only by the primary beam, but which still fits well within $3'\times 3'$. This confirms the COMAP pointing accuracy.

\end{appendix}

\end{document}